\documentclass[sigconf]{acmart}

\usepackage{booktabs}
\usepackage{multirow}
\usepackage{float}
\usepackage{placeins}
\usepackage{algorithm}
\usepackage{algorithmic}
\usepackage{colortbl}
\usepackage{xcolor}
\usepackage{amsthm}

\definecolor{bandgray}{gray}{0.93}

\newtheorem{theorem}{Theorem}

\setcopyright{acmlicensed}
\copyrightyear{2025}
\acmYear{2025}
\acmDOI{XXXXXXX.XXXXXXX}

\acmConference[KDD '25]{Proceedings of the 31st ACM SIGKDD Conference on Knowledge Discovery and Data Mining}{August 3--7, 2025}{Toronto, Canada}
\acmISBN{978-1-4503-XXXX-X/2025/08}

\begin{document}

\title{PEST: Physics-Enhanced Swin Transformer for 3D Turbulence Simulation}

\author{Yilong Dai}
\affiliation{%
   \institution{University of Alabama}
 \city{Tuscaloosa}
 \country{USA}}
 \email{ydai17@ua.edu}

\author{Shengyu Chen}
\affiliation{%
  \institution{University of Pittsburgh}
   \city{Pittsburgh}
  \country{USA}}
\email{shc160@pitt.edu}

\author{Xiaowei Jia}
\affiliation{%
  \institution{University of Pittsburgh}
     \city{Pittsburgh}
  \country{USA}}
\email{xiaowei@pitt.edu}

\author{Peyman Givi}
\affiliation{%
   \institution{University of Pittsburgh}
 \city{Pittsburgh}
 \country{USA}}
\email{peg10@pitt.edu}

\author{Runlong Yu}
\authornote{Runlong Yu is the corresponding author.}
\affiliation{%
   \institution{University of Alabama}
 \city{Tuscaloosa}
 \country{USA}}
\email{ryu5@ua.edu}
\renewcommand{\shortauthors}{Dai et al.}

\begin{abstract}
Accurate simulation of turbulent flows is fundamental to scientific and engineering applications. Direct numerical simulation (DNS) offers the highest fidelity but is computationally prohibitive, while existing data-driven alternatives struggle with stable long-horizon rollouts, physical consistency, and faithful simulation of small-scale structures. These challenges are particularly acute in three-dimensional (3D) settings, where the cubic growth of spatial degrees of freedom dramatically amplifies computational cost, memory demand, and the difficulty of capturing multi-scale interactions. To address these challenges, we propose a Physics-Enhanced Swin Transformer (PEST) for 3D turbulence simulation. PEST leverages a window-based self-attention mechanism to effectively model localized PDE interactions while maintaining computational efficiency. We introduce a frequency-domain adaptive loss that explicitly emphasizes small-scale structures, enabling more faithful simulation of high-frequency dynamics. To improve physical consistency, we incorporate Navier--Stokes residual constraints and divergence-free regularization directly into the learning objective. Extensive experiments on two representative turbulent flow configurations demonstrate that PEST achieves accurate, physically consistent, and stable autoregressive long-term simulations, outperforming existing data-driven baselines.
\end{abstract}




\settopmatter{printacmref=false} 
\renewcommand\footnotetextcopyrightpermission[1]{} 
\pagestyle{plain} 

\maketitle

\section{Introduction}
\label{sec:intro}
Multiscale partial differential equation (PDE) systems are ubiquitous in science and engineering, and their numerical solution is crucial for understanding complex physical phenomena and practical applications. Turbulence is one of the most classic and challenging examples of such systems, with its dynamics encompassing a cascade process from large-scale energy injection to small-scale energy dissipation~\cite{kolmogorov1941local,pope2000turbulent}. Accurate simulation of turbulence plays a critical role in aerospace~\cite{slotnick2014cfd}, energy engineering~\cite{veers2019grand}, and climate prediction~\cite{schneider2017earth}. Direct numerical simulation (DNS)~\cite{moin1998direct}, by directly solving the Navier-Stokes equations~\cite{temam2001navier} without introducing any turbulence modeling assumptions, can capture flow details at all scales with the highest numerical accuracy and physical fidelity. However, the computational cost of DNS increases dramatically with the Reynolds number~\cite{pope2000turbulent}, often making it infeasible for large-scale engineering turbulence problems. To address this limitation, data-driven methods have emerged as an alternative approach to accelerate turbulence simulation.

Data-driven methods have shown great potential in learning fluid dynamics from simulation data~\cite{dai2026learningpdesolversphysics}, with representative approaches spanning neural operators~\cite{li2021fno,wen2022ufno,kossaifi2023tfno}, attention-based architectures~\cite{wu2024transolver,li2023factformer,hao2024dpot}, physics-informed networks~\cite{raissi2019pinn,li2021pino}, and generative models~\cite{du2024confild}. However, applying these methods to high-resolution 3D turbulence simulations to completely replace DNS still faces multiple challenges. First, the spatiotemporal scale of 3D turbulence data presents significant computational complexity issues: the combination of high-resolution grids, multiple physical quantity channels, and long time series leads to dramatic computational and memory overhead, placing stringent demands on the efficiency and scalability of the architecture design. Second, the multi-scale nature of turbulence poses difficulties for model learning: because energy is mainly concentrated in large-scale structures, low-energy small-scale features are easily overshadowed during training. While this neglect has limited impact on short-term prediction accuracy, small-scale features are crucial for turbulent energy dissipation, and their absence can disrupt the physical integrity of the energy cascade and affect the stability of long-term simulations. Third, adherence to physical laws is a fundamental requirement for the reliability and trustworthiness of turbulence simulations: the flow field evolution must satisfy basic physical laws such as the Navier-Stokes equations and the divergence-free constraint for incompressible flows. Data-driven methods learn statistical patterns from data without explicit physical inductive biases and thus cannot guarantee satisfaction of these constraints without explicit physical guidance.

Existing methods attempt to address these challenges from different perspectives, but typically face trade-offs among the three aspects. Frequency-domain methods~\cite{li2021fno,wen2022ufno,kossaifi2023tfno} achieve resolution-invariance but lose the spatial locality that PDEs rely on, hindering effective imposition of physical constraints; some also sacrifice small-scale high-frequency information through frequency truncation to improve computational efficiency. Physical-space methods~\cite{wu2024transolver,li2023factformer,hao2024dpot} preserve spatial locality but struggle to balance computational efficiency with long-range dependency modeling, and typically rely on $\ell_2$ optimization, which inherently favors large-scale, high-energy features. Physics-informed approaches~\cite{raissi2019pinn,li2021pino,wang2021pideeponet} explicitly enforce governing equations but face challenges in balancing data and physics losses~\cite{wang2021understanding}, and their effectiveness depends heavily on the underlying architecture's ability to represent multi-scale dynamics. Generative methods~\cite{du2024confild} avoid autoregressive prediction and thus can mitigate error accumulation in long-term predictions, but varying noise levels make effective imposition of physical constraints challenging. These limitations motivate a unified approach that balances multi-scale feature learning and physical consistency while maintaining computational efficiency.

To address these challenges, this paper proposes the Physics-Enhanced Swin Transformer (PEST) for long-term simulation of high-resolution 3D turbulence. For architecture design, we employ the Swin Transformer as the backbone: its windowed attention mechanism effectively alleviates computational overhead in high-resolution scenarios while naturally aligning with the locality of PDEs, enabling effective learning of physical relationships between adjacent points in the flow field. The shifted window mechanism compensates for the limited receptive field of fixed windows, achieving a balance between local information and long-range dependencies. Furthermore, we introduce a gradient loss to further optimize information transfer at window boundaries. To address the challenge of multi-scale learning, we leverage Parseval's theorem to equivalently transform the data loss from the spatial domain to the frequency domain and design an adaptive weighting strategy to differentially optimize different frequency components, thereby enhancing the learning of low-energy small-scale features while preserving spatial locality. To address the challenge of physical constraints, we explicitly introduce adaptive Navier-Stokes equation residuals and divergence constraints into the loss function to enhance the physical consistency of predictions.

We validate the effectiveness of PEST on two representative turbulent configurations with distinct physical characteristics. Across both settings, PEST achieves stable autoregressive rollouts over extended horizons while consistently outperforming nine representative baselines. Beyond simulation accuracy, we evaluate physical consistency through multiple complementary metrics, confirming that PEST predictions remain physically plausible throughout long-term simulation. Ablation studies further verify the contribution of each proposed component.

In summary, our main contributions are as follows.

\begin{itemize}
    \item We propose PEST, a Physics-Enhanced Swin Transformer that unifies efficient multi-scale spatial modeling, frequency-adaptive spectral learning, and physics-informed regularization for stable, accurate 3D turbulence simulation.
    \item We introduce a frequency-adaptive spectral loss grounded in Parseval's theorem that rebalances learning across frequency bands via curriculum-guided adaptive weighting, addressing the inherent bias of $\ell_2$ loss toward large-scale dominance.
    \item We design an uncertainty-based multi-loss balancing strategy that jointly adapts all loss weights throughout training, resolving the scale mismatch between data-driven and physics-informed objectives.
    \item Extensive experiments on two turbulent flow benchmarks show that PEST consistently outperforms nine baselines in both simulation accuracy and physical consistency over extended autoregressive horizons.
\end{itemize}

\filbreak

\section{Related Work}
\label{sec:related_work}

Neural network-based PDE solvers have emerged as alternatives to traditional numerical methods. The Fourier Neural Operator~\cite{li2021fno} learns operator mappings in spectral space, with variants extending to multiscale~\cite{wen2022ufno} and spherical~\cite{guibas2022sfno} domains. Convolutional architectures like TF-Net~\cite{wang2020tfnet} operate in physical space for turbulent flow prediction. Transformer-based approaches have gained prominence: Transolver~\cite{wu2024transolver} and OFormer~\cite{li2023oformer,hao2023gnot} leverage attention for operator learning, while Swin Transformer~\cite{liu2021swin} addresses quadratic complexity through windowed computation and has been applied to weather forecasting~\cite{bi2023pangu,chen2023fengwu} and fluid simulation~\cite{11078485}. PDE-Transformer~\cite{janny2025pdetransformer,hao2024icon} further scales windowed attention to multi-PDE foundation models. Generative methods such as CoNFiLD~\cite{du2024confild} apply latent diffusion for turbulence generation. Our work adopts Swin Transformer, leveraging the alignment between windowed attention and local differential operators in PDEs.

Physics-Informed Neural Networks~\cite{raissi2019pinn,ren2024pmamba} incorporate PDE residuals as soft constraints. Extensions include domain decomposition methods such as XPINNs~\cite{jagtap2020xpinns} and cPINNs~\cite{jagtap2020cpinn}, as well as variational formulations like hp-VPINNs~\cite{kharazmi2021hpvpinn}. A persistent challenge is balancing data and physics losses: \citet{wang2021understanding} identified gradient flow pathologies and proposed rebalancing strategies, extended by~\citet{bischof2025relobralo} via multi-objective optimization. For time-dependent PDEs, \citet{wang2022causal} showed that standard PINNs can violate temporal causality and introduced causal training to enforce sequential learning, enabling the first successful PINN simulations of turbulent Navier--Stokes flows. For incompressible flows, divergence-free constraints have been enforced via vector potentials~\cite{richardson2024divergence} and specialized architectures~\cite{jin2021nsfnets}. Building on these developments, PEST addresses loss balancing through uncertainty-based adaptive weighting~\cite{kendall2018multi} for Navier--Stokes residuals, divergence constraints, and spectral data losses, within a Swin-based architecture whose windowed attention aligns with the local differential operators governing fluid motion.

A key challenge in data-driven turbulence modeling is that standard $\ell_2$ training inherently favors energy-dominant large scales, causing progressive loss of small-scale structures critical for long-term stability~\cite{stachenfeld2022learned,kochkov2021machine}. Spectral objectives~\cite{li2021pino,11078485} mitigate this but typically operate entirely in Fourier space, sacrificing compatibility with physical-space constraints. Multi-scale architectures such as U-Net/spectral hybrids~\cite{wen2022ufno} and factorized attention~\cite{li2023factformer} improve high-frequency preservation through structural design but still rely on scale-agnostic loss functions. FreqMoE~\cite{chen2025freqmoe} mitigates this via frequency-domain mixture-of-experts. Curriculum and adaptive weighting schemes~\cite{krishnapriyan2021characterizing,bischof2025relobralo} stabilize multi-objective optimization, while post-hoc refinement methods correct accumulated errors via denoising~\cite{lippe2024pderefiner} or low-resolution supervision~\cite{chen2024srtr}. DINO~\cite{wu2025dino} improves long-term rollout stability through operator decomposition into differential and integral branches. However, these efforts address multi-scale fidelity and physical consistency largely in isolation. PEST bridges this gap by coupling a frequency-adaptive spectral loss, grounded in Parseval's theorem to preserve spatial locality, with explicit Navier--Stokes and divergence constraints under uncertainty-based adaptive weighting.

\section{Problem Formulation}
\label{sec:problem}

We consider the simulation of 3D incompressible turbulent flows governed by the Navier-Stokes equations:
\begin{equation}
\frac{\partial \mathbf{u}}{\partial t} + (\mathbf{u} \cdot \nabla)\mathbf{u} = -\nabla p + \nu \nabla^2 \mathbf{u}, \quad \nabla \cdot \mathbf{u} = 0,
\label{eq:ns}
\end{equation}
where $\mathbf{u} = (u, v, w) \in \mathbb{R}^3$ denotes the velocity field, $p$ denotes the pressure field, and $\nu$ is the kinematic viscosity. The first equation describes momentum conservation, and the divergence-free condition $\nabla \cdot \mathbf{u} = 0$ enforces mass conservation for incompressible flows.

Let $\mathbf{s}^t = (\mathbf{u}^t, p^t) \in \mathbb{R}^{H \times W \times D \times 4}$ denote the flow state at time step $t$ on a 3D grid of spatial resolution $H \times W \times D$, consisting of three velocity components and pressure. We use $\hat{\mathbf{s}}$ and $\hat{\mathbf{u}}$ to denote predicted states and velocities, respectively. Given a sequence of $T$ consecutive flow states $\{\mathbf{s}^{t-T+1}, \ldots, \mathbf{s}^{t}\}$, our goal is to learn a mapping $\mathcal{F}_\theta$ that predicts the subsequent $T$ time steps:
\begin{equation}
\{\hat{\mathbf{s}}^{t+1}, \ldots, \hat{\mathbf{s}}^{t+T}\} = \mathcal{F}_\theta(\mathbf{s}^{t-T+1}, \ldots, \mathbf{s}^{t}),
\label{eq:task}
\end{equation}
where $\theta$ denotes the learnable parameters. In this work, we set $T=5$.

For long-term simulation, we apply $\mathcal{F}_\theta$ autoregressively: the predicted states $\{\hat{\mathbf{s}}^{t+1}, \ldots, \hat{\mathbf{s}}^{t+T}\}$ serve as input for the next prediction cycle, enabling simulation over extended time horizons.

The learning objective involves minimizing a combination of data loss and physics-based constraints. The data loss measures the discrepancy between predictions and the DNS reference solution. The physics-based constraints penalize violations of the Navier-Stokes equations and the divergence-free condition in Eq.~\ref{eq:ns}, ensuring that predictions remain physically consistent. The specific formulations and adaptive weighting strategies are detailed in the following subsections.


\begin{figure*}[t]
\centering
\includegraphics[width=\textwidth]{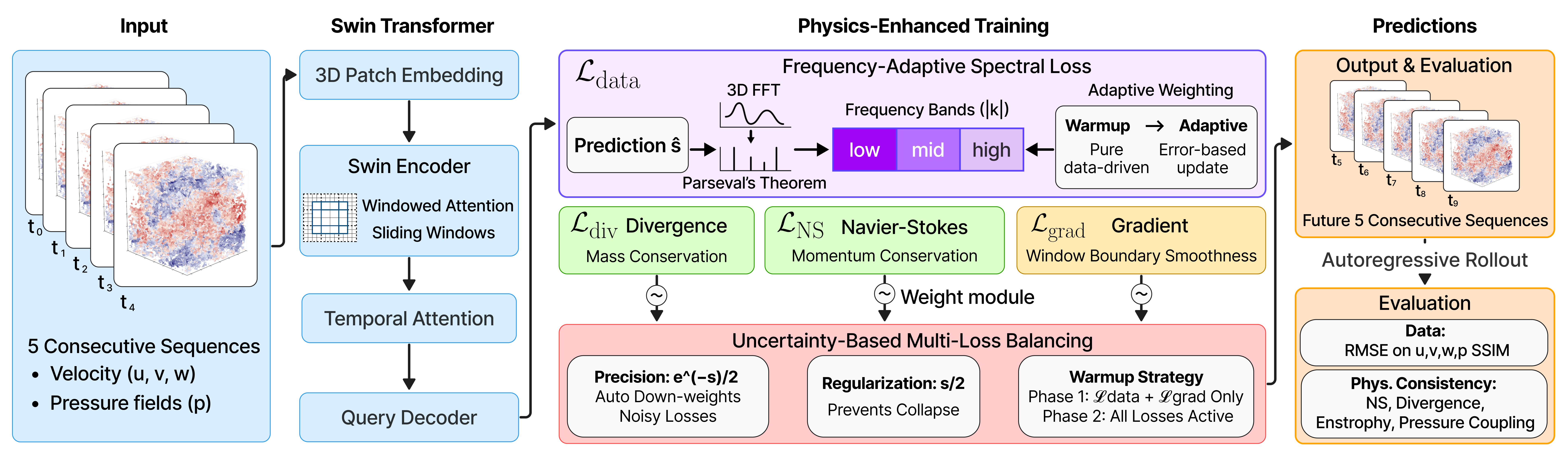}
\caption{
\textbf{Overview of the PEST framework.} The model takes 5 consecutive flow states as input, processes them through a Swin Transformer encoder--decoder, and predicts the next 5 timesteps. Training combines frequency-adaptive spectral loss, physics constraints, and gradient smoothness, balanced via uncertainty-based weighting.
}
\Description{Architecture diagram of PEST showing the input flow states, Swin Transformer encoder with windowed attention, decoder with learnable queries, and output predictions with multiple loss functions.}
\label{fig:architecture}
\end{figure*}

\section{Method}
\label{sec:method}

We present the PEST framework, whose overall architecture is illustrated in Fig.~\ref{fig:architecture}. PEST consists of four tightly integrated components. First, a 3D Swin Transformer backbone (\S\ref{sec:architecture}) processes high-resolution volumetric flow fields through windowed self-attention, achieving linear computational scaling while preserving the spatial locality inherent to PDE differential operators; a gradient smoothness loss is introduced to eliminate window-boundary artifacts. Second, a frequency-adaptive spectral loss (\S\ref{sec:spectral_loss}) leverages Parseval's theorem to equivalently reformulate the data simulation objective in Fourier space and applies curriculum-guided adaptive weighting across frequency bands, ensuring balanced learning of both energy-dominant large scales and physically critical small-scale structures. Third, physics-informed constraints (\S\ref{sec:physics_loss}) explicitly penalize violations of the Navier--Stokes equations and the divergence-free condition, anchoring predictions to the governing physics. Finally, an uncertainty-based multi-loss balancing mechanism (\S\ref{sec:uncertainty}) automatically adapts the relative weights of all loss terms throughout training, resolving scale mismatches and preventing any single objective from dominating optimization.

\subsection{Model Architecture}
\label{sec:architecture}
Turbulence simulation demands models that efficiently capture both local spatial interactions and long-range temporal dependencies across high-resolution 3D volumes. This challenge motivates our adoption of 3D Swin Transformer~\cite{liu2021swin}, which reduces attention complexity from $O(N^2)$ to $O(NM^3)$ through windowed computation while preserving the locality inherent to PDE differential operators, where $N$ denotes the total number of spatial tokens and $M$ is the window size.

\paragraph{Swin Transformer Backbone.}
The input sequence $\{\mathbf{s}_{t-T+1}, \ldots, \mathbf{s}_t\} \in \mathbb{R}^{T \times 4 \times D \times H \times W}$ is projected by a 3D convolutional patch embedding with stride $P$, yielding features $\mathbf{Z}^0 \in \mathbb{R}^{T \times C \times \frac{D}{P} \times \frac{H}{P} \times \frac{W}{P}}$, where $C$ is the embedding dimension. The encoder partitions the 3D volume into non-overlapping windows of size $M^3$ and computes self-attention within each window. Consecutive blocks alternate between regular and shifted windows (offset by $\frac{M}{2}$) to enable cross-window connectivity:
\begin{equation}
\mathbf{Z}^{l} = \text{FFN}\big(\text{(S)W-MSA}(\mathbf{Z}^{l-1})\big) + \mathbf{Z}^{l-1},
\end{equation}
where (S)W-MSA denotes (shifted) window multi-head self-attention and FFN is a feed-forward network. $\mathbf{Z}^{l}$ represents the feature tensor at layer $l$. A temporal attention module aggregates information across input frames with sinusoidal positional encodings. The decoder generates $T$ future predictions using learnable query tokens $\{\mathbf{q}_i\}_{i=1}^{T}$, where each query attends to encoder features via cross-attention and to other queries via temporal self-attention. A transposed 3D convolution recovers the original resolution, producing $\{\hat{\mathbf{s}}_{t+1}, \ldots, \hat{\mathbf{s}}_{t+T}\}$.

\paragraph{Gradient Smoothness Loss.}
While shifted windows enable cross-window information flow, we observe that window-based attention can still introduce discontinuities at window boundaries when flow structures span multiple windows (see Appendix~\ref{app:window_artifact}). To mitigate this artifact, we include a gradient matching loss:
\begin{equation}
\mathcal{L}_{\text{grad}} = \|\nabla \hat{\mathbf{s}} - \nabla \mathbf{s}\|_1,
\end{equation}
that encourages smooth predictions across the entire domain.

\subsection{Frequency-Adaptive Spectral Loss}
\label{sec:spectral_loss}

A fundamental challenge in turbulence modeling is the multi-scale nature of the flow: energy is predominantly concentrated in large-scale structures following the Kolmogorov $k^{-5/3}$ spectrum (where $k$ denotes the wavenumber magnitude), while small-scale features, though energetically subdominant, are crucial for turbulent dissipation and maintaining the physical integrity of the energy cascade. Standard $\ell_2$ loss inherently biases optimization toward high-energy low-frequency components, causing the model to neglect fine-scale structures essential for long-term simulation stability. In this section, we introduce a frequency-adaptive spectral loss that explicitly reweights contributions across scales to ensure balanced multi-scale learning.

Our approach is grounded in the following classical result from harmonic analysis:

\begin{theorem}[Parseval's Theorem]
\label{thm:parseval}
For any square-integrable function $f$ defined on a discrete domain of $N$ points and its discrete Fourier transform $\mathcal{F}(f)$, the total energy is preserved:
\begin{equation}
\|f\|_2^2 = \frac{1}{N}\|\mathcal{F}(f)\|_2^2.
\label{eq:parseval_thm}
\end{equation}
\end{theorem}

Applying Theorem~\ref{thm:parseval} to the prediction residual $\hat{\mathbf{s}} - \mathbf{s}$ yields:
\begin{equation}
\|\hat{\mathbf{s}} - \mathbf{s}\|_2^2 = \frac{1}{N}\|\mathcal{F}(\hat{\mathbf{s}}) - \mathcal{F}(\mathbf{s})\|_2^2,
\label{eq:parseval}
\end{equation}
where $\mathcal{F}$ denotes the 3D discrete Fourier transform. This identity establishes a precise equivalence between spatial-domain and frequency-domain error measurement, which is the theoretical foundation of our loss design. Crucially, it implies that \emph{the standard $\ell_2$ loss in physical space can be exactly decomposed into per-wavenumber contributions in Fourier space}. This decomposition enables us to selectively reweight individual frequency bands without altering the fundamental nature of the loss---the total remains a valid spatial simulation error, merely with non-uniform emphasis across scales. Unlike purely spectral methods that operate entirely in frequency domain (e.g., spectral convolutions in FNO~\cite{li2021fno}), our formulation preserves full compatibility with spatial differential operators required for computing divergence and Navier--Stokes residuals in physical space (\S\ref{sec:physics_loss}). In other words, Parseval's theorem allows us to enjoy the \emph{scale-aware controllability} of the frequency domain while retaining the \emph{spatial locality} needed for physics-informed constraints---a combination that neither purely spatial nor purely spectral approaches can achieve alone.

We partition the frequency domain into three bands $\mathcal{B}$ = \{low, mid, high\} based on wavenumber magnitude $|\mathbf{k}|$, where $\mathbf{k} = (k_x, k_y, k_z)$ is the 3D wavenumber vector. The spectral loss is defined as:
\begin{equation}
\mathcal{L}_{\text{data}} = \sum_{b \in \mathcal{B}} w_b \cdot \frac{1}{|S_b|} \sum_{\mathbf{k} \in S_b} |\mathcal{F}(\hat{\mathbf{s}})_{\mathbf{k}} - \mathcal{F}(\mathbf{s})_{\mathbf{k}}|^2,
\label{eq:spectral_loss}
\end{equation}
where $S_b$ denotes the set of wavenumbers in band $b$, and $w_b$ are learnable band weights that adapt during training to balance simulation accuracy across scales.
We employ a curriculum strategy that initially prioritizes large-scale structures during warmup, then gradually shifts emphasis toward high-frequency features. In the adaptive phase, weights are dynamically adjusted based on per-band simulation errors:
\begin{equation}
w_b \leftarrow w_b + \eta \left( \frac{\sqrt{\epsilon_b}}{\sum_{b'} \sqrt{\epsilon_{b'}}} - \frac{w_b}{\sum_{b'} w_{b'}} \right),
\label{eq:adaptive_weight}
\end{equation}
where $\epsilon_b$ is the exponential moving average of the simulation error for band $b$, and $\eta$ is the adaptation rate. This mechanism automatically allocates more weight to frequency bands with higher errors, promoting balanced learning across scales.

\subsection{Physics-Informed Constraints}
\label{sec:physics_loss}

While data-driven losses optimize for simulation accuracy, they do not guarantee that predictions obey the underlying physical laws governing fluid motion. We incorporate two physics-based loss terms that enforce fundamental conservation principles, computed in physical space after denormalization to ensure correct units.

\paragraph{Divergence Constraint.}
For incompressible flows, mass conservation requires the velocity field to be divergence-free: $\nabla \cdot \mathbf{u} = 0$. Violating this constraint leads to unphysical mass sources or sinks in the predicted flow field. We enforce incompressibility through:
\begin{equation}
\mathcal{L}_{\text{div}} = \|\nabla \cdot \hat{\mathbf{u}}\|_2^2,
\end{equation}
where the divergence is computed using central finite differences. Note that while the original high-resolution DNS data satisfies $\nabla \cdot \mathbf{u} = 0$, spatial downsampling to lower resolutions introduces aliasing artifacts that can result in non-zero divergence when computed on the coarser grid. To handle such data, we use a residual formulation $\mathcal{L}_{\text{div}} = \|\nabla \cdot \hat{\mathbf{u}} - \nabla \cdot \mathbf{u}\|_2^2$ that matches the divergence of predictions to that of the DNS solution rather than enforcing zero divergence directly, where the divergence residual is computed independently for each predicted time step.

\paragraph{Navier-Stokes Residual.}
The Navier-Stokes equations describe momentum conservation in viscous fluids, governing how velocity evolves under pressure gradients, advection, and viscous diffusion. We penalize deviations from this fundamental law by computing the residual:
\begin{equation}
\mathcal{L}_{\text{NS}} = \left\| \frac{\partial \mathbf{u}}{\partial t} + (\mathbf{u} \cdot \nabla)\mathbf{u} + \nabla p - \nu \nabla^2 \mathbf{u} \right\|_2^2,
\end{equation}
where spatial derivatives use central finite differences and the temporal derivative is approximated via central differences between consecutive predicted frames, falling back to forward differences when boundary frames are involved. The residual is evaluated at a single representative time step within each prediction window to balance physical regularization with computational cost. This soft constraint encourages the model to learn dynamics consistent with the governing equations rather than merely fitting training data.

\subsection{Uncertainty-Based Multi-Loss Balancing}
\label{sec:uncertainty}

Training PEST involves multiple loss terms with vastly different scales and dynamics. At the beginning of training, when predictions deviate significantly from physical solutions, physics-based losses ($\mathcal{L}_{\text{div}}$, $\mathcal{L}_{\text{NS}}$) can be several orders of magnitude larger than the data simulation loss. Naively combining these losses causes physics constraints to dominate optimization, overwhelming the gradient signal from $\mathcal{L}_{\text{data}}$ and leading to slow convergence or training instability. Moreover, the relative magnitudes evolve during training: as predictions become more physically consistent, physics losses decrease rapidly while spectral losses on high-frequency components may remain challenging to reduce. Fixed weighting schemes cannot adapt to these dynamics.

To address this, we adopt the homoscedastic uncertainty framework~\cite{kendall2018multi} to automatically learn optimal loss weights. For each loss term $\mathcal{L}_i$, we introduce a learnable parameter $s_i = \log \sigma_i^2$ representing task uncertainty. The total loss becomes:
\begin{equation}
\mathcal{L}_{\text{total}} = \sum_{i} \left( \frac{1}{2} e^{-s_i} \mathcal{L}_i + \frac{1}{2} s_i \right).
\end{equation}
The precision term $e^{-s_i}/2$ acts as an adaptive weight: when a loss is large or noisy, the corresponding $\sigma_i$ increases, automatically down-weighting its contribution to prevent gradient explosion. The regularization term $s_i/2$ prevents degenerate solutions where all weights collapse to zero. The parameters $\{s_i\}$ are optimized jointly with model parameters using a separate optimizer.

To ensure stable optimization, we apply a warmup strategy: during early training, we train with only $\mathcal{L}_{\text{data}}$ and $\mathcal{L}_{\text{grad}}$ while physics losses remain inactive. This allows the model to first establish basic prediction capability before physics constraints are introduced and uncertainty parameters begin adapting.


\section{Experiments}
\label{sec:experiments}

We conduct experiments on two representative 3D turbulent flow benchmarks to evaluate PEST. Specifically, we seek to answer three research questions:
\begin{itemize}
    \item \textbf{RQ1}: Can PEST achieve accurate and stable predictions over extended autoregressive horizons compared to existing methods?
    \item \textbf{RQ2}: Do PEST's predictions remain physically consistent with the governing Navier--Stokes equations?
    \item \textbf{RQ3}: How does each proposed component contribute to the overall performance?
\end{itemize}

\subsection{Datasets}
\label{sec:datasets}

We consider two turbulent flow configurations with distinct physical characteristics. Each flow state comprises four field components: three velocity components $u$, $v$, $w$ and the pressure field $p$.

\paragraph{JHU Isotropic Turbulence.}
The Johns Hopkins Turbulence Database~\cite{li2008jhtdb} provides pseudo-spectral DNS solution data of forced isotropic turbulence on a $1024^3$ periodic grid. We use a spatially downsampled $128^3$ subset at Taylor-scale Reynolds number $Re_\lambda \approx 433$ ($\nu = 1.85 \times 10^{-4}$), with 200 frames at temporal resolution $\delta t = 0.01$. This dataset represents statistically stationary turbulence with a fully developed energy cascade.

\paragraph{Taylor-Green Vortex.}
The Taylor-Green vortex~\cite{taylor1937mechanism} is a canonical benchmark for turbulence transition and decay, solved using a high-order finite-difference DNS at $Re = 1600$ on a $64 \times 128 \times 128$ grid ($\delta t = 2$). Unlike the stationary JHU dataset, this configuration captures transient dynamics across laminar-to-turbulent transition and subsequent decay.

\paragraph{Data Splitting.}
We employ chunk-based temporal partitioning rather than a simple chronological cutoff. A naive split that reserves only the final frames for testing would bias evaluation toward a single flow regime, which is especially problematic for the Taylor-Green vortex where dynamics evolve rapidly through distinct stages (transition, peak dissipation, decay). Instead, the timeline is divided into non-overlapping chunks; within each chunk, earlier frames form the training set and later frames are reserved for validation and testing, with temporal gaps between splits. This interleaved design ensures that test samples span diverse flow regimes while preventing any temporal leakage from training data. Each sample consists of 5 input frames and 5 target frames.

\subsection{Experimental Setup}
\label{sec:setup}

\paragraph{Implementation.}
All models are constrained to fit on a single GPU. For PEST, we apply patch embedding to reduce GPU memory cost. Frequency modes are truncated accordingly to maintain computational feasibility. The Swin encoder has three stages with depths $[2, 8, 2]$ and embedding dimension $C = 512$; the total model size is approximately 170M parameters. We train for 100 epochs with AdamW (learning rate $3 \times 10^{-5}$) using mixed-precision on a single NVIDIA 5090 GPU (32GB). Full hyperparameters are provided in Appendix~\ref{app:exp_details}.

\paragraph{Baselines.}
We compare against nine baselines spanning spectral methods (FNO3D~\cite{li2021fno}, U-FNO~\cite{wen2022ufno}, TFNO~\cite{kossaifi2023tfno}), attention-based architectures (Transolver~\cite{wu2024transolver}, DPOT~\cite{hao2024dpot}, FactFormer~\cite{li2023factformer}), and physics-informed approaches (PINO~\cite{li2021pino}, DeepONet~\cite{lu2021deeponet}, PI-DeepONet~\cite{wang2021pideeponet}). Baseline descriptions are provided in Appendix~\ref{app:exp_details}.

\paragraph{Metrics.}
We evaluate using \textit{Root Mean Square Error (RMSE)}, defined as the normalized $\ell_2$ error $\|\hat{\mathbf{s}} - \mathbf{s}\|_2 / \|\mathbf{s}\|_2$, and \textit{Structural Similarity Index (SSIM)}~\cite{wang2004ssim}, which captures the preservation of local spatial patterns beyond pointwise error. For autoregressive evaluation, JHU supports 3 rollout rounds (R1--R3, frames 1--15), while Taylor-Green supports 2 rounds (R1--R2).

\subsection{Main Results}
\label{sec:main_results}

\begin{table*}[t]
\caption{Autoregressive prediction on JHU Isotropic Turbulence. RMSE (physical space) and SSIM across three rollout rounds. Best in \textbf{bold}, second best \underline{underlined}.}
\label{tab:jhu_main}
\setlength{\tabcolsep}{2.2pt}
\fontsize{8pt}{10pt}\selectfont
\begin{tabular}{l cccccc cccccc cccccc cccccc}
\toprule
& \multicolumn{6}{c}{$u$} & \multicolumn{6}{c}{\cellcolor{bandgray}$v$} & \multicolumn{6}{c}{$w$} & \multicolumn{6}{c}{\cellcolor{bandgray}$p$} \\
\cmidrule(lr){2-7} \cmidrule(lr){8-13} \cmidrule(lr){14-19} \cmidrule(lr){20-25}
& \multicolumn{3}{c}{RMSE $\downarrow$} & \multicolumn{3}{c}{SSIM $\uparrow$} & \multicolumn{3}{c}{\cellcolor{bandgray}RMSE $\downarrow$} & \multicolumn{3}{c}{\cellcolor{bandgray}SSIM $\uparrow$} & \multicolumn{3}{c}{RMSE $\downarrow$} & \multicolumn{3}{c}{SSIM $\uparrow$} & \multicolumn{3}{c}{\cellcolor{bandgray}RMSE $\downarrow$} & \multicolumn{3}{c}{\cellcolor{bandgray}SSIM $\uparrow$} \\
Method & R1 & R2 & R3 & R1 & R2 & R3 & \cellcolor{bandgray}R1 & \cellcolor{bandgray}R2 & \cellcolor{bandgray}R3 & \cellcolor{bandgray}R1 & \cellcolor{bandgray}R2 & \cellcolor{bandgray}R3 & R1 & R2 & R3 & R1 & R2 & R3 & \cellcolor{bandgray}R1 & \cellcolor{bandgray}R2 & \cellcolor{bandgray}R3 & \cellcolor{bandgray}R1 & \cellcolor{bandgray}R2 & \cellcolor{bandgray}R3 \\
\midrule
FNO3D & .163 & .234 & .265 & .625 & .484 & .415 & \cellcolor{bandgray}.161 & \cellcolor{bandgray}.234 & \cellcolor{bandgray}.265 & \cellcolor{bandgray}.690 & \cellcolor{bandgray}.521 & \cellcolor{bandgray}.445 & .162 & .236 & .269 & .633 & .465 & .394 & \cellcolor{bandgray}.086 & \cellcolor{bandgray}.121 & \cellcolor{bandgray}.141 & \cellcolor{bandgray}.599 & \cellcolor{bandgray}.512 & \cellcolor{bandgray}.474 \\
U-FNO & \underline{.153} & .219 & .259 & .659 & .525 & .437 & \cellcolor{bandgray}\underline{.150} & \cellcolor{bandgray}.216 & \cellcolor{bandgray}.255 & \cellcolor{bandgray}.717 & \cellcolor{bandgray}.572 & \cellcolor{bandgray}.482 & \underline{.153} & .219 & .256 & .645 & .511 & .429 & \cellcolor{bandgray}\underline{.083} & \cellcolor{bandgray}.118 & \cellcolor{bandgray}.139 & \cellcolor{bandgray}.632 & \cellcolor{bandgray}.532 & \cellcolor{bandgray}.482 \\
TFNO & .177 & .283 & .344 & .628 & .425 & .314 & \cellcolor{bandgray}.174 & \cellcolor{bandgray}.284 & \cellcolor{bandgray}.349 & \cellcolor{bandgray}.686 & \cellcolor{bandgray}.446 & \cellcolor{bandgray}.326 & .180 & .286 & .351 & .610 & .398 & .289 & \cellcolor{bandgray}.091 & \cellcolor{bandgray}.157 & \cellcolor{bandgray}.205 & \cellcolor{bandgray}.582 & \cellcolor{bandgray}.472 & \cellcolor{bandgray}.395 \\
Transolver & .211 & .231 & .257 & .469 & .536 & .424 & \cellcolor{bandgray}.213 & \cellcolor{bandgray}.232 & \cellcolor{bandgray}.258 & \cellcolor{bandgray}.459 & \cellcolor{bandgray}.521 & \cellcolor{bandgray}.424 & .212 & .232 & .258 & .447 & .504 & .404 & \cellcolor{bandgray}.117 & \cellcolor{bandgray}.128 & \cellcolor{bandgray}.147 & \cellcolor{bandgray}.617 & \cellcolor{bandgray}.684 & \cellcolor{bandgray}.591 \\
DPOT & .203 & .226 & .253 & .658 & .577 & .498 & \cellcolor{bandgray}.205 & \cellcolor{bandgray}.228 & \cellcolor{bandgray}.257 & \cellcolor{bandgray}.629 & \cellcolor{bandgray}.557 & \cellcolor{bandgray}.487 & .204 & .228 & .256 & .618 & .541 & .468 & \cellcolor{bandgray}.112 & \cellcolor{bandgray}.125 & \cellcolor{bandgray}.146 & \cellcolor{bandgray}.764 & \cellcolor{bandgray}.705 & \cellcolor{bandgray}.637 \\
FactFormer & .171 & \underline{.213} & \underline{.252} & \underline{.755} & \underline{.626} & \underline{.508} & \cellcolor{bandgray}.171 & \cellcolor{bandgray}\underline{.213} & \cellcolor{bandgray}\underline{.252} & \cellcolor{bandgray}\underline{.734} & \cellcolor{bandgray}\underline{.611} & \cellcolor{bandgray}\underline{.503} & .171 & \underline{.213} & \underline{.251} & \underline{.729} & \underline{.600} & \underline{.487} & \cellcolor{bandgray}.092 & \cellcolor{bandgray}\underline{.114} & \cellcolor{bandgray}\underline{.139} & \cellcolor{bandgray}\underline{.820} & \cellcolor{bandgray}\underline{.734} & \cellcolor{bandgray}\underline{.649} \\
PINO & .163 & .234 & .265 & .630 & .485 & .415 & \cellcolor{bandgray}.162 & \cellcolor{bandgray}.235 & \cellcolor{bandgray}.267 & \cellcolor{bandgray}.678 & \cellcolor{bandgray}.509 & \cellcolor{bandgray}.432 & .163 & .236 & .268 & .628 & .469 & .400 & \cellcolor{bandgray}.088 & \cellcolor{bandgray}.121 & \cellcolor{bandgray}\underline{.139} & \cellcolor{bandgray}.574 & \cellcolor{bandgray}.503 & \cellcolor{bandgray}.472 \\
DeepONet & .293 & .298 & .302 & .368 & .359 & .351 & \cellcolor{bandgray}.297 & \cellcolor{bandgray}.302 & \cellcolor{bandgray}.307 & \cellcolor{bandgray}.367 & \cellcolor{bandgray}.359 & \cellcolor{bandgray}.354 & .295 & .300 & .305 & .347 & .340 & .334 & \cellcolor{bandgray}.154 & \cellcolor{bandgray}.160 & \cellcolor{bandgray}.165 & \cellcolor{bandgray}.609 & \cellcolor{bandgray}.595 & \cellcolor{bandgray}.584 \\
PI-DeepONet & .298 & .301 & .305 & .350 & .343 & .337 & \cellcolor{bandgray}.304 & \cellcolor{bandgray}.307 & \cellcolor{bandgray}.311 & \cellcolor{bandgray}.350 & \cellcolor{bandgray}.345 & \cellcolor{bandgray}.340 & .302 & .305 & .309 & .331 & .326 & .322 & \cellcolor{bandgray}.154 & \cellcolor{bandgray}.158 & \cellcolor{bandgray}.162 & \cellcolor{bandgray}.607 & \cellcolor{bandgray}.597 & \cellcolor{bandgray}.587 \\
\midrule
\textbf{PEST} & \textbf{.127} & \textbf{.165} & \textbf{.191} & \textbf{.878} & \textbf{.796} & \textbf{.745} & \cellcolor{bandgray}\textbf{.128} & \cellcolor{bandgray}\textbf{.174} & \cellcolor{bandgray}\textbf{.208} & \cellcolor{bandgray}\textbf{.880} & \cellcolor{bandgray}\textbf{.779} & \cellcolor{bandgray}\textbf{.715} & \textbf{.128} & \textbf{.173} & \textbf{.206} & \textbf{.873} & \textbf{.774} & \textbf{.708} & \cellcolor{bandgray}\textbf{.060} & \cellcolor{bandgray}\textbf{.086} & \cellcolor{bandgray}\textbf{.104} & \cellcolor{bandgray}\textbf{.922} & \cellcolor{bandgray}\textbf{.905} & \cellcolor{bandgray}\textbf{.881} \\
\bottomrule
\end{tabular}
\end{table*}

\begin{table*}[!h]
\caption{Autoregressive prediction on DNS Taylor-Green Vortex. RMSE (physical space) and SSIM across two rollout rounds. Best in \textbf{bold}, second best \underline{underlined}.}
\label{tab:dns_main}
\setlength{\tabcolsep}{2.8pt}
\fontsize{8pt}{10pt}\selectfont
\begin{tabular}{l cccc cccc cccc cccc}
\toprule
& \multicolumn{4}{c}{$u$} & \multicolumn{4}{c}{\cellcolor{bandgray}$v$} & \multicolumn{4}{c}{$w$} & \multicolumn{4}{c}{\cellcolor{bandgray}$p$} \\
\cmidrule(lr){2-5} \cmidrule(lr){6-9} \cmidrule(lr){10-13} \cmidrule(lr){14-17}
& \multicolumn{2}{c}{RMSE $\downarrow$} & \multicolumn{2}{c}{SSIM $\uparrow$} & \multicolumn{2}{c}{\cellcolor{bandgray}RMSE $\downarrow$} & \multicolumn{2}{c}{\cellcolor{bandgray}SSIM $\uparrow$} & \multicolumn{2}{c}{RMSE $\downarrow$} & \multicolumn{2}{c}{SSIM $\uparrow$} & \multicolumn{2}{c}{\cellcolor{bandgray}RMSE $\downarrow$} & \multicolumn{2}{c}{\cellcolor{bandgray}SSIM $\uparrow$} \\
Method & R1 & R2 & R1 & R2 & \cellcolor{bandgray}R1 & \cellcolor{bandgray}R2 & \cellcolor{bandgray}R1 & \cellcolor{bandgray}R2 & R1 & R2 & R1 & R2 & \cellcolor{bandgray}R1 & \cellcolor{bandgray}R2 & \cellcolor{bandgray}R1 & \cellcolor{bandgray}R2 \\
\midrule
FNO3D & .040 & .066 & .830 & .685 & \cellcolor{bandgray}.038 & \cellcolor{bandgray}.066 & \cellcolor{bandgray}.853 & \cellcolor{bandgray}.695 & .065 & .073 & .531 & .454 & \cellcolor{bandgray}.062 & \cellcolor{bandgray}.062 & \cellcolor{bandgray}.461 & \cellcolor{bandgray}.385 \\
U-FNO & .038 & .061 & .850 & .708 & \cellcolor{bandgray}.038 & \cellcolor{bandgray}.062 & \cellcolor{bandgray}.851 & \cellcolor{bandgray}.696 & .065 & .076 & .541 & .420 & \cellcolor{bandgray}.054 & \cellcolor{bandgray}.056 & \cellcolor{bandgray}.534 & \cellcolor{bandgray}.424 \\
TFNO & .036 & .086 & .872 & .632 & \cellcolor{bandgray}.037 & \cellcolor{bandgray}.085 & \cellcolor{bandgray}.866 & \cellcolor{bandgray}.637 & .050 & .077 & .666 & .414 & \cellcolor{bandgray}.036 & \cellcolor{bandgray}.048 & \cellcolor{bandgray}.661 & \cellcolor{bandgray}.462 \\
Transolver & .036 & .051 & .883 & \underline{.790} & \cellcolor{bandgray}.036 & \cellcolor{bandgray}\underline{.051} & \cellcolor{bandgray}.884 & \cellcolor{bandgray}\underline{.791} & .045 & .058 & \underline{.767} & \underline{.664} & \cellcolor{bandgray}\underline{.021} & \cellcolor{bandgray}\underline{.026} & \cellcolor{bandgray}\underline{.809} & \cellcolor{bandgray}\underline{.700} \\
DPOT & .038 & .055 & .867 & .766 & \cellcolor{bandgray}.038 & \cellcolor{bandgray}.055 & \cellcolor{bandgray}.867 & \cellcolor{bandgray}.768 & .046 & .056 & .762 & .658 & \cellcolor{bandgray}.022 & \cellcolor{bandgray}.027 & \cellcolor{bandgray}.790 & \cellcolor{bandgray}.660 \\
FactFormer & \underline{.031} & \underline{.049} & \underline{.892} & .787 & \cellcolor{bandgray}\underline{.031} & \cellcolor{bandgray}\underline{.049} & \cellcolor{bandgray}\underline{.893} & \cellcolor{bandgray}.787 & \underline{.040} & \underline{.055} & \textbf{.826} & \underline{.687} & \cellcolor{bandgray}\underline{.020} & \cellcolor{bandgray}\underline{.025} & \cellcolor{bandgray}.788 & \cellcolor{bandgray}\underline{.670} \\
PINO & .041 & .064 & .831 & .694 & \cellcolor{bandgray}.041 & \cellcolor{bandgray}.066 & \cellcolor{bandgray}.821 & \cellcolor{bandgray}.681 & .068 & .077 & .543 & .421 & \cellcolor{bandgray}.060 & \cellcolor{bandgray}.060 & \cellcolor{bandgray}.473 & \cellcolor{bandgray}.399 \\
DeepONet & .068 & .072 & .651 & .623 & \cellcolor{bandgray}.071 & \cellcolor{bandgray}.074 & \cellcolor{bandgray}.636 & \cellcolor{bandgray}.611 & .067 & .073 & .528 & .491 & \cellcolor{bandgray}.028 & \cellcolor{bandgray}.031 & \cellcolor{bandgray}.668 & \cellcolor{bandgray}.615 \\
PI-DeepONet & .075 & .085 & .616 & .574 & \cellcolor{bandgray}.080 & \cellcolor{bandgray}.089 & \cellcolor{bandgray}.609 & \cellcolor{bandgray}.568 & .068 & .065 & .472 & .485 & \cellcolor{bandgray}.030 & \cellcolor{bandgray}.037 & \cellcolor{bandgray}.571 & \cellcolor{bandgray}.483 \\
\midrule
\textbf{PEST} & \textbf{.026} & \textbf{.036} & \textbf{.951} & \textbf{.909} & \cellcolor{bandgray}\textbf{.026} & \cellcolor{bandgray}\textbf{.036} & \cellcolor{bandgray}\textbf{.951} & \cellcolor{bandgray}\textbf{.908} & \textbf{.024} & \textbf{.026} & .764 & \textbf{.743} & \cellcolor{bandgray}\textbf{.010} & \cellcolor{bandgray}\textbf{.013} & \cellcolor{bandgray}\textbf{.970} & \cellcolor{bandgray}\textbf{.942} \\
\bottomrule
\end{tabular}
\end{table*}

\begin{figure*}[t]
\centering
\includegraphics[width=\textwidth]{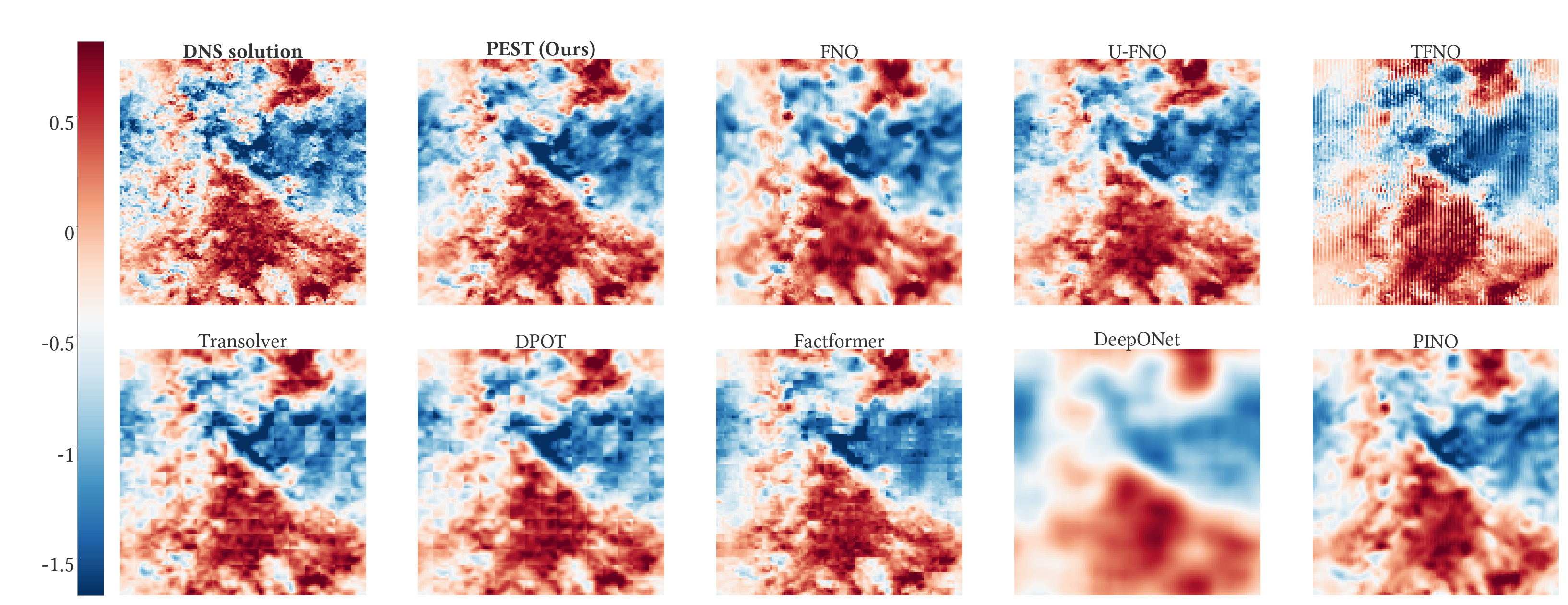}
\caption{
\textbf{$w$-component predictions on JHU at autoregressive Round 3 ($t=6.0$).} Color bar indicates physical-space velocity in m/s. PEST closely matches the DNS solution; baselines show varying degrees of over-smoothing or structural degradation.
}
\Description{Visual comparison of w-velocity component predictions from different methods on JHU turbulence dataset at Round 3. PEST shows the closest match to DNS solution with preserved fine-scale structures, while baselines show varying degrees of smoothing and structural degradation.}
\label{fig:jhu_w_round3_comparison}
\end{figure*}

\begin{figure}[t]
\centering
\includegraphics[width=\columnwidth]{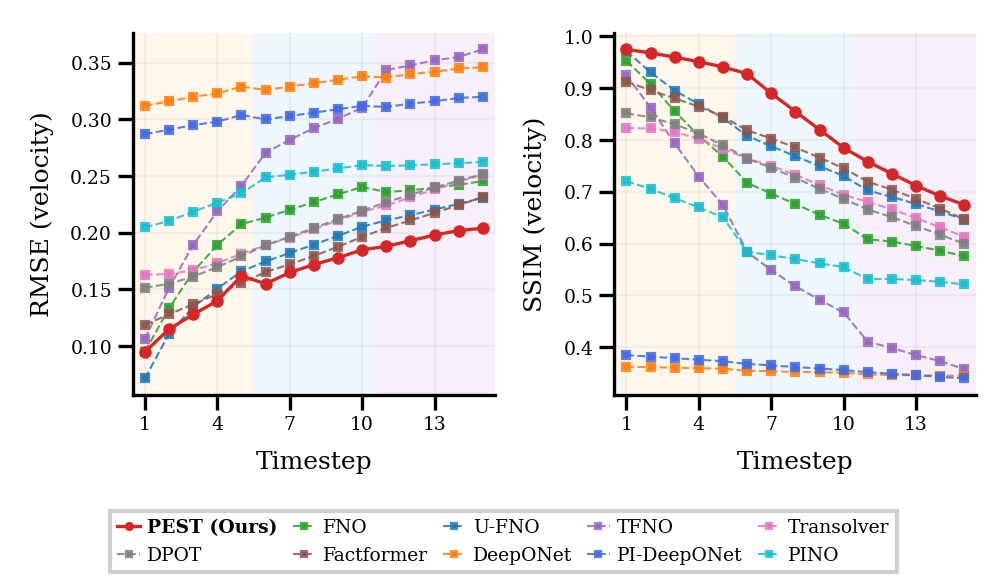}
\caption{\textbf{Per-timestep RMSE and SSIM on JHU across 15 autoregressive steps (3 rounds).}}
\Description{Line plots showing per-timestep RMSE increasing and SSIM decreasing over 15 autoregressive steps for PEST and baselines. PEST maintains consistently lower RMSE and higher SSIM throughout all three rounds.}
\label{fig:jhu_rmse_ssim_comparison}
\end{figure}

\paragraph{RQ1: Prediction Accuracy and Stability.}
Each model predicts 5 future frames from 5 input frames, and autoregressive rollout proceeds in 5-step rounds rather than single-step advances. Table~\ref{tab:jhu_main} presents the autoregressive prediction results on the JHU dataset. PEST consistently outperforms all baselines across rollout rounds and velocity components, achieving 17\% lower RMSE than the second-best method (U-FNO) in Round 1 and maintaining this advantage through Round 3. Notably, PEST attains an SSIM above 0.87 for all velocity components in Round 1 and still exceeds 0.70 at Round 3, whereas the best baseline (FactFormer) drops to approximately 0.50, indicating that PEST preserves fine-scale turbulent structures far more effectively over extended rollouts. Fig.~\ref{fig:jhu_rmse_ssim_comparison} shows per-timestep RMSE and SSIM across the 15-step prediction horizon, where the advantage becomes more pronounced in later rounds and the stable SSIM trajectory confirms preservation of fine-scale structures throughout long-term rollouts. We present a representative qualitative comparison in Fig.~\ref{fig:jhu_w_round3_comparison}: the $w$-velocity component at autoregressive Round 3 ($t=6.0$). Additional visualizations for all components and rounds are provided in Appendix~\ref{app:visualizations}.

Table~\ref{tab:dns_main} shows results on the Taylor-Green dataset with 2 rollout rounds. The transient nature of this flow, spanning laminar-to-turbulent transition and subsequent decay, presents additional challenges, yet PEST maintains superior performance with 16\% lower RMSE than FactFormer in Round 1. The improvement is especially pronounced in the pressure field, where PEST achieves an RMSE of 0.01 versus 0.02 for the second-best method, and an SSIM of 0.97, indicating near-perfect pressure simulation. Furthermore, PEST's advantage is well preserved from Round 1 to Round 2 across all components, suggesting that physics-informed training provides effective regularization against error accumulation even in transient flow regimes. Per-timestep analysis (Appendix~\ref{app:physics_analysis}) confirms that PEST maintains SSIM above 0.9 for $u$, $v$, and $p$ components throughout the 10-step prediction horizon.

\begin{figure*}[h]
\centering
\includegraphics[width=\textwidth]{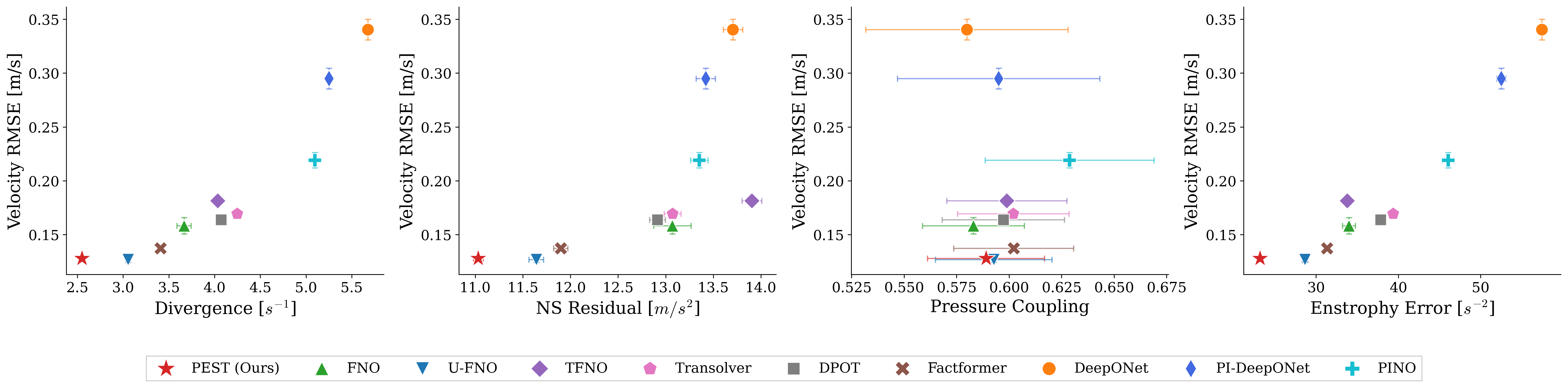}
\caption{
\textbf{Physical consistency analysis on JHU (Round 1).} Velocity RMSE vs.\ four physics metrics; lower-left is better. PEST achieves the best trade-off across all metrics.
}
\Description{Scatter plots showing physics consistency metrics versus velocity RMSE on JHU dataset for Round 1. PEST (red star) is positioned at the lower-left corner in all four subplots, indicating superior performance in both prediction accuracy and physical consistency compared to all baseline methods.}
\label{fig:jhu_physics_r1}
\end{figure*}

\vspace{-6pt}
\paragraph{RQ2: Physical Consistency.}
Beyond simulation accuracy, we evaluate whether PEST predictions obey the governing physics using four complementary metrics (Fig.~\ref{fig:jhu_physics_r1}). The first two correspond to our training losses: \textit{divergence residual} measuring violation of mass conservation ($\nabla \cdot \mathbf{u} = 0$), and \textit{Navier-Stokes residual} measuring deviation from momentum conservation. We additionally evaluate two metrics \textit{not used during training}: \textit{pressure coupling error}, which measures consistency between the predicted pressure and the pressure implied by the velocity field via the Poisson equation $\nabla^2 p = -\rho \nabla \mathbf{u} : \nabla \mathbf{u}^\top$; and \textit{enstrophy error}, which measures the accuracy of vorticity $\boldsymbol{\omega} = \nabla \times \mathbf{u}$ through enstrophy $\mathcal{E} = \frac{1}{2}|\boldsymbol{\omega}|^2$. Detailed definitions are provided in Appendix~\ref{app:physics_analysis}.

For Round 1 predictions on JHU, PEST achieves the best trade-off between prediction accuracy and physical consistency, positioned at the lower-left corner across all four metrics. This advantage persists through subsequent rollout rounds (see Appendix~\ref{app:physics_analysis}), demonstrating that our physics-informed constraints provide sustained regularization rather than diminishing with error accumulation. Notably, the improvement on the two unseen metrics (pressure coupling and enstrophy) confirms that PEST learns genuinely physical dynamics rather than merely minimizing the penalized losses.

\subsection{Ablation Studies (RQ3)}
\label{sec:ablation}

\paragraph{Component Analysis.}
Table~\ref{tab:ablation_rmse} and~\ref{tab:ablation_ssim} present a systematic ablation on the JHU dataset. Starting from a vanilla Swin Transformer baseline, we incrementally add each proposed component: (1) gradient smoothness loss (\textbf{+Grad.}), (2) frequency-adaptive spectral loss and physics constraints (\textbf{+Phy.}), and (3) the full PEST model with uncertainty-based multi-loss balancing (\textbf{PEST (Ours)}). Each component contributes to improved performance, and their benefits compound across rollout rounds.

\begin{table}[h]
\caption{Ablation study on JHU: physical RMSE ($\downarrow$).}
\label{tab:ablation_rmse}
\setlength{\tabcolsep}{2pt}
\footnotesize
\begin{tabular}{l ccc ccc ccc ccc}
\toprule
& \multicolumn{3}{c}{$u$} & \multicolumn{3}{c}{\cellcolor{bandgray}$v$} & \multicolumn{3}{c}{$w$} & \multicolumn{3}{c}{\cellcolor{bandgray}$p$} \\
Method & R1 & R2 & R3 & \cellcolor{bandgray}R1 & \cellcolor{bandgray}R2 & \cellcolor{bandgray}R3 & R1 & R2 & R3 & \cellcolor{bandgray}R1 & \cellcolor{bandgray}R2 & \cellcolor{bandgray}R3 \\
\midrule
Vanilla Swin & .183 & .204 & .228 & \cellcolor{bandgray}.189 & \cellcolor{bandgray}.210 & \cellcolor{bandgray}.234 & .186 & .207 & .231 & \cellcolor{bandgray}.091 & \cellcolor{bandgray}.106 & \cellcolor{bandgray}.123 \\
+ Grad. & .139 & .181 & .219 & \cellcolor{bandgray}.146 & \cellcolor{bandgray}.188 & \cellcolor{bandgray}.229 & .141 & .201 & .225 & \cellcolor{bandgray}.064 & \cellcolor{bandgray}.091 & \cellcolor{bandgray}.112 \\
+ Phy. & .129 & .169 & .201 & \cellcolor{bandgray}.130 & \cellcolor{bandgray}.185 & \cellcolor{bandgray}.215 & .131 & .185 & .211 & \cellcolor{bandgray}.062 & \cellcolor{bandgray}.089 & \cellcolor{bandgray}.106 \\
\midrule
\textbf{PEST (Ours)} & \textbf{.127} & \textbf{.165} & \textbf{.191} & \cellcolor{bandgray}\textbf{.128} & \cellcolor{bandgray}\textbf{.174} & \cellcolor{bandgray}\textbf{.208} & \textbf{.128} & \textbf{.173} & \textbf{.206} & \cellcolor{bandgray}\textbf{.060} & \cellcolor{bandgray}\textbf{.086} & \cellcolor{bandgray}\textbf{.104} \\
\bottomrule
\end{tabular}
\end{table}

\begin{table}[h]
\caption{Ablation study on JHU: SSIM ($\uparrow$).}
\label{tab:ablation_ssim}
\setlength{\tabcolsep}{2pt}
\footnotesize
\begin{tabular}{l ccc ccc ccc ccc}
\toprule
& \multicolumn{3}{c}{$u$} & \multicolumn{3}{c}{\cellcolor{bandgray}$v$} & \multicolumn{3}{c}{$w$} & \multicolumn{3}{c}{\cellcolor{bandgray}$p$} \\
Method & R1 & R2 & R3 & \cellcolor{bandgray}R1 & \cellcolor{bandgray}R2 & \cellcolor{bandgray}R3 & R1 & R2 & R3 & \cellcolor{bandgray}R1 & \cellcolor{bandgray}R2 & \cellcolor{bandgray}R3 \\
\midrule
Vanilla Swin & .797 & .753 & .702 & \cellcolor{bandgray}.780 & \cellcolor{bandgray}.740 & \cellcolor{bandgray}.693 & .773 & .729 & .678 & \cellcolor{bandgray}.913 & \cellcolor{bandgray}.845 & \cellcolor{bandgray}.791 \\
+ Grad. & .849 & .772 & .721 & \cellcolor{bandgray}.853 & \cellcolor{bandgray}.756 & \cellcolor{bandgray}.698 & .845 & .754 & .681 & \cellcolor{bandgray}.917 & \cellcolor{bandgray}.889 & \cellcolor{bandgray}.863 \\
+ Phy. & .867 & .782 & .734 & \cellcolor{bandgray}.861 & \cellcolor{bandgray}.766 & \cellcolor{bandgray}.701 & .859 & .764 & .698 & \cellcolor{bandgray}.919 & \cellcolor{bandgray}.898 & \cellcolor{bandgray}.872 \\
\midrule
\textbf{PEST (Ours)} & \textbf{.878} & \textbf{.796} & \textbf{.745} & \cellcolor{bandgray}\textbf{.880} & \cellcolor{bandgray}\textbf{.779} & \cellcolor{bandgray}\textbf{.715} & \textbf{.873} & \textbf{.774} & \textbf{.708} & \cellcolor{bandgray}\textbf{.922} & \cellcolor{bandgray}\textbf{.905} & \cellcolor{bandgray}\textbf{.881} \\
\bottomrule
\end{tabular}
\end{table}

The gradient loss alone yields the largest single-component improvement, reducing Round 1 velocity RMSE by 24\% (averaged across $u$, $v$, $w$) and eliminating the window boundary artifacts discussed in Appendix~\ref{app:window_artifact}. Adding physics constraints and the spectral loss further reduces RMSE by approximately 8\%, with gains more pronounced in later rounds (R3 improvement of 6\% over +Grad.), confirming that physics-informed regularization is especially beneficial for long-term stability. The full PEST model with uncertainty-based balancing provides an additional consistent improvement over fixed-weight physics constraints, demonstrating that adaptive loss weighting enables more effective optimization when multiple loss terms with different scales are combined.

\begin{figure}[t]
\centering
\includegraphics[width=0.70\columnwidth]{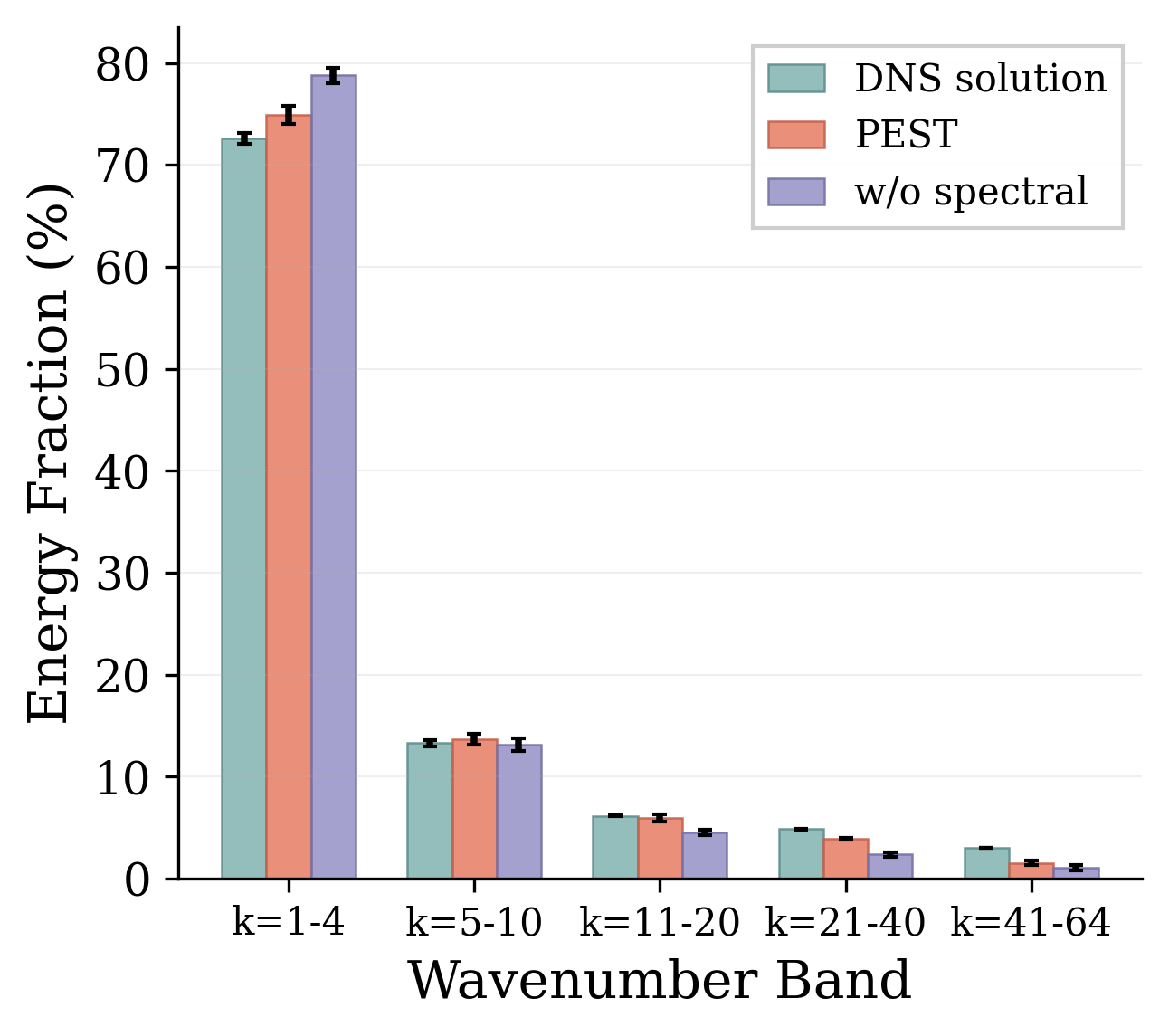}
\caption{\textbf{Spectral ablation: kinetic energy across frequency bands.}}
\vspace{-0.3cm}
\Description{Bar chart comparing kinetic energy preservation across low, mid, and high frequency bands with and without frequency-adaptive spectral loss. Adaptive weighting improves agreement with DNS across all bands except k=5--10.}
\label{fig:spectral_ablation}
\end{figure}

\paragraph{Spectral Loss Analysis.}
Fig.~\ref{fig:spectral_ablation} compares the kinetic energy spectrum with and without the frequency-adaptive spectral loss. The adaptive weighting improves energy preservation across multiple scales: low-frequency ($k=1$--$4$), mid-to-high-frequency ($k=11$--$40$), and high-frequency ($k>40$) bands all show improved agreement with the DNS solution, while only $k=5$--$10$ remains comparable. This confirms that our adaptive weighting successfully balances multi-scale simulation beyond just the energy-dominant large scales. A detailed frequency-resolved analysis (Appendix~\ref{app:spectral_ablation}) further reveals that the improvement spans a broad range of individual wavenumbers, validating the physical motivation of emphasizing under-represented scales.

\FloatBarrier

\section{Conclusion}
\label{sec:conclusion}

We presented PEST, a Physics-Enhanced Swin Transformer for autoregressive simulation of 3D turbulence. PEST addresses three key challenges facing data-driven turbulence simulation through complementary design choices: a windowed attention architecture that aligns with PDE locality while maintaining computational efficiency, a frequency-adaptive spectral loss that rebalances multi-scale learning to preserve energetically subdominant but physically important small-scale structures, and explicit physics constraints with uncertainty-based adaptive weighting that enforce Navier-Stokes consistency without destabilizing training. Experiments on two turbulent flow configurations with distinct physical characteristics demonstrate that PEST consistently outperforms nine baselines in both simulation accuracy and physical consistency over extended autoregressive rollouts. This work takes a step toward reliable neural surrogate models for turbulence, and we hope the design principles that combine spatial locality with spectral reweighting and adaptive physics regularization can prove useful for other multi-scale PDE systems.

\section{Interdisciplinary Collaborations}
\label{sec:interdisciplinary}

This work arises from collaboration between machine learning researchers (University of Alabama) and domain experts in computational fluid dynamics (University of Pittsburgh), including Peyman Givi, who has over three decades of research in turbulence simulation and modeling. This partnership ensures that PEST integrates ML architecture design with established physical theory and governing equations throughout its framework. From the ML side, key challenges include scaling transformers to 3D volumes and stabilizing multi-objective optimization across loss terms differing by orders of magnitude. From the domain side, the central contribution is a neural surrogate that respects conservation laws over extended horizons, a property critical for downstream scientific analysis.

\section{Limitations and Ethical Considerations}
\label{sec:limitations}

Both datasets are publicly available simulation products that contain no personal or sensitive information. We acknowledge several limitations. First, our evaluation uses structured-grid DNS data with complete spatiotemporal coverage; real-world turbulent flows from experiments or unstructured meshes may introduce irregular distributions and sensor noise not captured here. Second, computational constraints limit training to batch size 1 on a single GPU at resolutions up to $128^3$; scaling to higher resolutions or Reynolds numbers may require architectural adaptations. Third, while PEST shows strong autoregressive stability, prediction quality inevitably degrades over very long horizons; combining neural predictions with post-hoc refinement or occasional numerical solver corrections is a promising direction for further extending simulation fidelity.
\FloatBarrier

\section{GenAI Disclosure}
\label{sec:genai}
We used Claude to help polish the English manuscript text. All research ideas, model designs, experiments, and scientific conclusions were conceived, implemented, and validated by the authors.

\bibliographystyle{ACM-Reference-Format}
\bibliography{references}

\appendix

\section{Overview of Supplementary Material}
\label{app:overview}

This appendix provides comprehensive supplementary material organized as follows.

\begin{itemize}
    \item \textbf{Appendix~\ref{app:exp_details}: Detailed Experimental Setup.} Complete dataset statistics, architecture specifications, training hyperparameters, and metric definitions that complement the experimental setup in \S\ref{sec:setup}.

    \item \textbf{Appendix~\ref{app:baselines}: Baseline Descriptions.} Concise descriptions of all nine baseline methods spanning spectral, attention-based, and physics-informed categories.

    \item \textbf{Appendix~\ref{app:window_artifact}: Window Boundary Artifacts.} Diagnostic analysis of the checkerboard artifacts produced by vanilla Swin Transformer on structured flows, with visual evidence that the gradient smoothness loss eliminates this issue.

    \item \textbf{Appendix~\ref{app:physics_analysis}: Extended Physical Consistency Analysis.} Definitions of all four physics metrics (divergence residual, Navier--Stokes residual, pressure coupling error, enstrophy error), extended scatter-plot results for JHU Rounds~2--3, and per-timestep analysis on Taylor-Green Vortex.

    \item \textbf{Appendix~\ref{app:spectral_ablation}: Detailed Spectral Ablation.} Fine-grained (11-band) frequency decomposition of kinetic energy preservation, extending the coarse five-band analysis in \S\ref{sec:ablation}.

    \item \textbf{Appendix~\ref{app:spectral_analysis}: Spectral Energy Characterization.} Intrinsic energy distributions of both datasets, explaining why the frequency-adaptive spectral loss yields larger gains on JHU than on Taylor-Green Vortex.

    \item \textbf{Appendix~\ref{app:plugin}: Plug-in Refinement Methods.} Evaluation of two orthogonal, model-agnostic refinement strategies---test-time low-resolution supervision and PDE-Refiner denoising---that further improve PEST's long-horizon performance.

    \item \textbf{Appendix~\ref{app:visualizations}: Additional Visualizations.} Complete qualitative comparisons for all field components, datasets, and rollout rounds, including full autoregressive rollout sequences.
\end{itemize}

\noindent\textbf{How to read the results.}
Tables report RMSE~($\downarrow$) and SSIM~($\uparrow$) per field component ($u$, $v$, $w$, $p$) and per rollout round (R1, R2, R3 for JHU; R1, R2 for Taylor-Green).
Scatter plots in the physics consistency analysis (Appendix~\ref{app:physics_analysis}) show velocity RMSE on the $y$-axis versus a physics metric on the $x$-axis; the ideal position is the \emph{lower-left} corner, indicating simultaneously low prediction error and high physical consistency.
Spectral analyses (Appendices~\ref{app:spectral_ablation}--\ref{app:spectral_analysis}) compare energy fractions across wavenumber bands; closer agreement with the DNS solution indicates better multi-scale fidelity.

\section{Detailed Experimental Setup}
\label{app:exp_details}

\subsection{Dataset Details}

\paragraph{JHU Isotropic Turbulence.}
The original simulation is performed on a $1024^3$ periodic grid using a pseudo-spectral code with domain size $(2\pi)^3$.
The flow is characterized by Taylor-scale Reynolds number $Re_\lambda \approx 433$, kinematic viscosity $\nu = 1.85 \times 10^{-4}$, and large-eddy turnover time $T_L \approx 2.0$.
The temporal resolution is $\delta t = 0.01$, with 200 frames spanning approximately one turnover time.

\paragraph{Taylor-Green Vortex.}
Starting from a smooth analytical initial condition, the flow undergoes vortex stretching and breakdown, reaching peak dissipation at $t/t_c \approx 8$ before decaying.
The DNS is performed at Reynolds number $Re = 1600$ ($\nu = 1/1600$) on a $64 \times 128 \times 128$ grid with temporal resolution $\delta t = 2$.

\paragraph{Data Splits.}
For both datasets, we employ chunk-based temporal splitting to prevent data leakage while ensuring coverage of diverse flow regimes.
The timeline is divided into non-overlapping chunks; within each chunk, earlier frames are used for training and later frames for validation/testing, with temporal gaps between splits.
For JHU, we use 5 chunks across 200 frames, yielding 80 training and 55 test samples.
For Taylor-Green, the chunk-based split ensures that test samples span different stages of the turbulence evolution (transition, peak dissipation, and decay).
Each sample consists of 5 input frames and 5 target frames.

\subsection{Architecture Details}

We use patch stride $P = (4, 4, 4)$ for JHU ($128^3$) and $P = (2, 4, 4)$ for Taylor-Green ($64 \times 128 \times 128$), yielding a uniform patched resolution of $32^3$ in both cases.
The Swin encoder has three stages with depths $[2, 8, 2]$ and attention heads $[8, 16, 32]$.
The embedding dimension is $C = 512$ with window size $M = 8$.
The decoder has 6 layers with dimension 384 and 12 heads.
The total model size is approximately 170M parameters.

\subsection{Training Details}

We use the AdamW optimizer with learning rate $3 \times 10^{-5}$, weight decay $0.01$, and a cosine schedule with 5-epoch warmup.
For uncertainty parameters, we use a separate Adam optimizer with learning rate $10^{-3}$.
The foundation phase (data-only training without physics losses) comprises the first 30\% of epochs.
We train for 100 epochs with batch size 1 using mixed-precision on a single NVIDIA 5090 GPU (32\,GB).

\paragraph{Spectral Loss Configuration.}
Frequency bands are partitioned at $k_{\max}/3$ and $2k_{\max}/3$, where $k_{\max}$ is the maximum resolvable wavenumber determined by the grid resolution.
Initial weights $(w_{\text{low}}, w_{\text{mid}}, w_{\text{high}}) = (2.0, 1.0, 0.5)$ transition to $(1.0, 1.5, 2.0)$ over the first 1000 steps, then adapt dynamically based on per-band errors with rate $\eta = 0.1$.

\paragraph{Metrics.}
\textit{RMSE} (Root Mean Square Error) measures the relative $\ell_2$ error: $\|\hat{\mathbf{s}} - \mathbf{s}\|_2 / \|\mathbf{s}\|_2$, computed in physical space after denormalization.
\textit{SSIM} (Structural Similarity Index Measure)~\cite{wang2004ssim} evaluates perceptual similarity by comparing local patterns of luminance, contrast, and structure, capturing the preservation of fine-scale spatial features that RMSE may overlook.

\section{Baseline Descriptions}
\label{app:baselines}

We compare PEST against nine representative neural PDE solvers spanning three categories.

\paragraph{Spectral Methods.}

\textit{FNO3D}~\cite{li2021fno} learns mappings between function spaces via spectral convolutions in Fourier domain.
The kernel is parameterized directly in frequency space, enabling efficient global receptive fields with linear complexity.
We extend the original 2D formulation to 3D volumetric data.

\textit{U-FNO}~\cite{wen2022ufno} combines U-Net's multi-scale encoder--decoder architecture with spectral convolutions.
Each encoder/decoder stage applies Fourier-domain filtering followed by spatial up/downsampling, with skip connections preserving high-frequency details across scales.

\textit{TFNO}~\cite{kossaifi2023tfno} (Tensorized FNO) improves FNO's parameter efficiency via Tucker decomposition of spectral weights, factorizing the full weight tensor into a compact core tensor and factor matrices.

\paragraph{Attention-Based Architectures.}

\textit{Transolver}~\cite{wu2024transolver} reduces quadratic attention complexity via physics-aware point-to-slice aggregation, compressing input points into $K$ learnable slice tokens ($K \ll N$) and broadcasting back to the original resolution to achieve $O(NK)$ complexity.

\textit{DPOT}~\cite{hao2024dpot} (Denoising Pretraining Operator Transformer) performs attention in Fourier space through learnable spectral filtering.
Each attention layer applies element-wise learnable weights to frequency coefficients.
We omit pre-training for fair comparison.

\textit{FactFormer}~\cite{li2023factformer} factorizes 3D attention into three sequential 1D axial attentions along spatial dimensions, reducing complexity from $O(N^2)$ to $O(N^{4/3})$ while maintaining a global receptive field.

\paragraph{Physics-Informed Approaches.}

\textit{PINO}~\cite{li2021pino} (Physics-Informed Neural Operator) augments FNO with physics-based regularization, combining the spectral convolution architecture with soft constraints enforcing incompressibility and momentum conservation.

\textit{DeepONet}~\cite{lu2021deeponet} learns operators via branch--trunk decomposition: the branch network encodes the input function into basis coefficients, while the trunk network learns spatial basis functions.

\textit{PI-DeepONet}~\cite{wang2021pideeponet} extends DeepONet with physics loss regularization, augmenting the branch--trunk architecture with soft constraints on divergence-free velocity fields and Navier--Stokes residuals.

\section{Window Boundary Artifacts in Swin Transformer}
\label{app:window_artifact}

This section diagnoses a systematic artifact that arises when vanilla Swin Transformer is applied to flows with large-scale coherent structures, and demonstrates that the gradient smoothness loss ($\mathcal{L}_{\text{grad}}$) proposed in \S\ref{sec:architecture} effectively eliminates it.

\paragraph{Origin of the Artifact.}
The Taylor-Green vortex evolves from a smooth analytical initial condition through vortex stretching toward turbulent breakdown.
In the early stages ($t/t_c < 4$), the flow is dominated by large-scale vortical structures whose spatial extent approaches or exceeds the window size used in Swin Transformer's local attention.
In standard Swin Transformer, each window computes self-attention independently before the shifted-window operation enables cross-window communication in the subsequent layer.
For large-scale structures spanning multiple windows, features near window edges receive attention from fewer relevant tokens than those at window centers, creating systematic prediction bias at window boundaries.

\begin{figure}[H]
\centering
\includegraphics[width=\columnwidth]{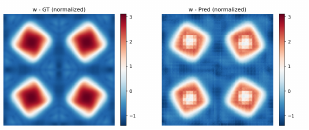}
\caption{
\textbf{Vanilla Swin Transformer without gradient loss.}
Left: DNS ground truth ($w$-component, normalized). Right: model prediction.
The predicted field exhibits checkerboard artifacts aligned with the $8 \times 8 \times 8$ window grid, manifesting as unphysical ``hollow cores'' in vortex regions.
}
\Description{Visualization showing checkerboard artifacts in predicted velocity field without gradient loss, appearing as hollow cores in vortex regions aligned with the window grid.}
\label{fig:window_artifact_bad}
\end{figure}

\begin{figure}[H]
\centering
\includegraphics[width=\columnwidth]{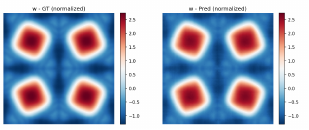}
\caption{
\textbf{Swin Transformer with gradient smoothness loss.}
Left: DNS ground truth ($w$-component, normalized). Right: model prediction.
The gradient loss eliminates window-boundary discontinuities, producing smooth predictions consistent with the DNS solution.
}
\Description{Visualization showing smooth predicted velocity field with gradient loss, matching the DNS solution without checkerboard artifacts.}
\label{fig:window_artifact_good}
\end{figure}

\paragraph{Visual Evidence.}
As shown in Fig.~\ref{fig:window_artifact_bad}, without gradient smoothness loss, the predicted velocity field exhibits a distinctive checkerboard pattern aligned with the $8 \times 8 \times 8$ window grid.
The artifact is most pronounced in vortex cores, appearing as an unphysical ``hollow core'' pattern where the predicted field intensity drops sharply at window boundaries.
Fig.~\ref{fig:window_artifact_good} shows that incorporating $\mathcal{L}_{\text{grad}}$ eliminates the artifact entirely, yielding smooth predictions that faithfully capture the large-scale vortical structures.

\paragraph{Window Size Comparison.}
To verify that the artifact stems from the attention mechanism itself rather than a particular window size, we compare two configurations on the vanilla Swin Transformer (Tables~\ref{tab:win_size_rmse_comparison} and~\ref{tab:win_size_ssim_comparison}).
The $(8,8,8)$ and $(4,4,4)$ windows yield nearly identical performance across all components and rollout rounds, confirming that the artifact is inherent to windowed attention and that the gradient loss addresses the root cause regardless of window configuration.

\begin{table}[t]
\caption{RMSE comparison of two window size configurations on vanilla Swin Transformer (JHU dataset, $\downarrow$). Performance is nearly identical, confirming the artifact is independent of window size.}
\label{tab:win_size_rmse_comparison}
\setlength{\tabcolsep}{2pt}
\footnotesize
\begin{tabular}{l ccc ccc ccc ccc}
\toprule
& \multicolumn{3}{c}{$u$} & \multicolumn{3}{c}{\cellcolor{bandgray}$v$} & \multicolumn{3}{c}{$w$} & \multicolumn{3}{c}{\cellcolor{bandgray}$p$} \\
Window Size  & R1 & R2 & R3 & \cellcolor{bandgray}R1 & \cellcolor{bandgray}R2 & \cellcolor{bandgray}R3 & R1 & R2 & R3 & \cellcolor{bandgray}R1 & \cellcolor{bandgray}R2 & \cellcolor{bandgray}R3 \\
\midrule
(8,8,8) & .183 & .204 & .228 & \cellcolor{bandgray}.189 & \cellcolor{bandgray}.210 & \cellcolor{bandgray}.234 & .186 & .207 & .231 & \cellcolor{bandgray}.091 & \cellcolor{bandgray}.106 & \cellcolor{bandgray}.123 \\
(4,4,4) & .185 & .205 & .227 & \cellcolor{bandgray}.190 & \cellcolor{bandgray}.211 & \cellcolor{bandgray}.235 & .188 & .209 & .232 & \cellcolor{bandgray}.092 & \cellcolor{bandgray}.106 & \cellcolor{bandgray}.123 \\
\bottomrule
\end{tabular}
\end{table}

\begin{table}[t]
\caption{SSIM comparison of two window size configurations on vanilla Swin Transformer (JHU dataset, $\uparrow$). Consistent with RMSE, the two configurations perform comparably.}
\label{tab:win_size_ssim_comparison}
\setlength{\tabcolsep}{2pt}
\footnotesize
\begin{tabular}{l ccc ccc ccc ccc}
\toprule
& \multicolumn{3}{c}{$u$} & \multicolumn{3}{c}{\cellcolor{bandgray}$v$} & \multicolumn{3}{c}{$w$} & \multicolumn{3}{c}{\cellcolor{bandgray}$p$} \\
Window Size & R1 & R2 & R3 & \cellcolor{bandgray}R1 & \cellcolor{bandgray}R2 & \cellcolor{bandgray}R3 & R1 & R2 & R3 & \cellcolor{bandgray}R1 & \cellcolor{bandgray}R2 & \cellcolor{bandgray}R3 \\
\midrule
(8,8,8) & .797 & .753 & .702 & \cellcolor{bandgray}.780 & \cellcolor{bandgray}.740 & \cellcolor{bandgray}.693 & .773 & .729 & .678 & \cellcolor{bandgray}.913 & \cellcolor{bandgray}.845 & \cellcolor{bandgray}.791 \\
(4,4,4) & .792 & .750 & .701 & \cellcolor{bandgray}.777 & \cellcolor{bandgray}.736 & \cellcolor{bandgray}.690 & .768 & .723 & .673 & \cellcolor{bandgray}.912 & \cellcolor{bandgray}.844 & \cellcolor{bandgray}.791 \\
\bottomrule
\end{tabular}
\end{table}

\paragraph{Takeaway.}
Windowed self-attention introduces systematic boundary artifacts on flows with large-scale coherent structures.
The gradient smoothness loss provides a simple and effective remedy that is agnostic to window size, enabling the Swin Transformer backbone to operate on structured flows without sacrificing spatial continuity.

\section{Extended Physical Consistency Analysis}
\label{app:physics_analysis}

This section complements the Round~1 physical consistency results presented in Fig.~\ref{fig:jhu_physics_r1} of the main text.
We first define all four physics metrics, then present extended results for JHU Rounds~2--3 and per-timestep analysis on the Taylor-Green Vortex.

\subsection{Physics Metrics Definitions}
\label{app:physics_metrics}

We evaluate physical consistency using four complementary metrics.
The first two (\textit{divergence residual} and \textit{Navier--Stokes residual}) correspond to terms in our training loss, while the remaining two (\textit{pressure coupling error} and \textit{enstrophy error}) are \emph{not} used during training and thus serve as independent diagnostics of whether the model has learned physically meaningful dynamics.

\paragraph{Divergence Residual (Mass Conservation).}
For incompressible flows, mass conservation requires $\nabla \cdot \mathbf{u} = 0$. We measure the root mean square (RMS) divergence residual:
\begin{equation}
\text{Div}_{\text{RMS}} = \sqrt{\frac{1}{N}\sum_i \left(\frac{\partial u}{\partial x} + \frac{\partial v}{\partial y} + \frac{\partial w}{\partial z}\right)_i^2},
\end{equation}
where spatial derivatives are computed using second-order central differences.
For downsampled data where aliasing may introduce non-zero DNS divergence, we use the residual form $\|\nabla \cdot \hat{\mathbf{u}} - \nabla \cdot \mathbf{u}\|_{\text{RMS}}$.

\paragraph{Navier--Stokes Residual (Momentum Conservation).}
The momentum equation residual measures deviation from the governing dynamics:
\begin{equation}
R_{\text{NS}} = \frac{\partial \mathbf{u}}{\partial t} + (\mathbf{u} \cdot \nabla)\mathbf{u} + \nabla p - \nu \nabla^2 \mathbf{u},
\end{equation}
where the temporal derivative uses central differences between consecutive frames, and spatial derivatives use second-order finite differences. We report $\|R_{\text{NS}}\|_{\text{RMS}}$ in units of m/s$^2$.

\paragraph{Pressure Coupling Error (Velocity--Pressure Consistency).}
The pressure field must satisfy the Poisson equation derived from taking the divergence of the momentum equation:
\begin{equation}
\nabla^2 p = -\rho \sum_{i,j} \frac{\partial u_i}{\partial x_j} \frac{\partial u_j}{\partial x_i}.
\end{equation}
We solve this equation spectrally using FFT to obtain $p_{\text{computed}}$ from the predicted velocity field, then measure $\|p_{\text{pred}} - p_{\text{computed}}\|_{\text{RMS}}$.
This metric verifies internal consistency between predicted velocity and pressure \emph{without} being directly optimized during training.

\paragraph{Enstrophy Error (Vorticity Accuracy).}
Enstrophy $\mathcal{E} = \frac{1}{2}|\boldsymbol{\omega}|^2$ quantifies rotational energy, where vorticity $\boldsymbol{\omega} = \nabla \times \mathbf{u}$.
For turbulent flows, accurate vorticity prediction is crucial as it characterizes the energy cascade from large to small scales.
We report the RMS enstrophy error between predictions and the DNS solution.

\subsection{Extended JHU Physical Consistency Results}
\label{app:jhu_physics_extended}

\begin{figure*}[t]
\centering
\includegraphics[width=\textwidth]{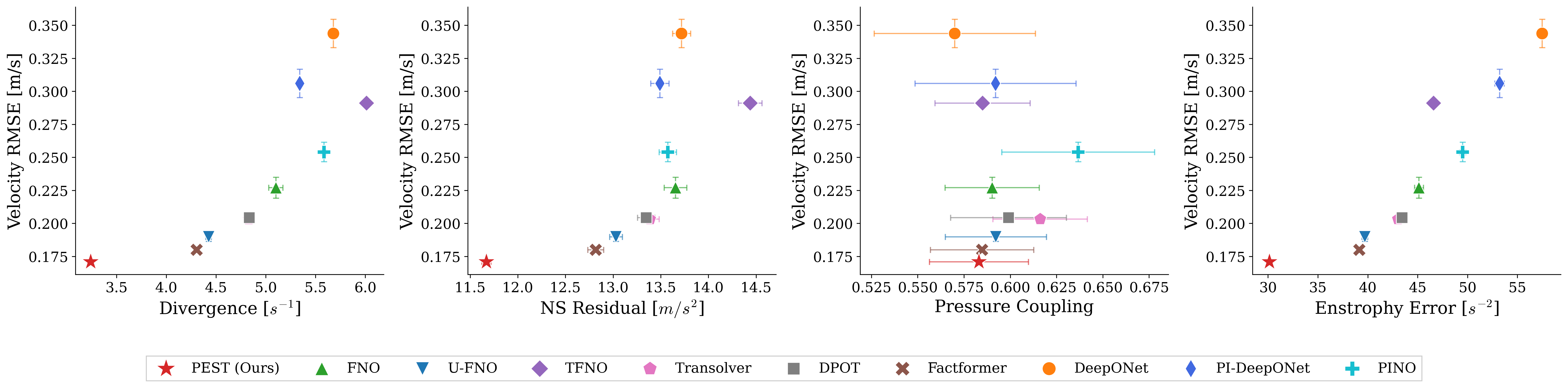}
\caption{
\textbf{Physical consistency on JHU (Round~2).}
Each panel plots velocity RMSE ($y$-axis) against one of the four physics metrics ($x$-axis); the lower-left corner is optimal.
Despite accumulated autoregressive errors, PEST (red star) maintains the best trade-off across all four metrics, with relatively small error bars indicating robust predictions across different flow realizations.
}
\Description{Scatter plots showing physics consistency metrics versus velocity RMSE on JHU dataset for Round 2. PEST remains at the lower-left corner across all four metrics.}
\label{fig:jhu_physics_r2}
\end{figure*}

\begin{figure*}[t]
\centering
\includegraphics[width=\textwidth]{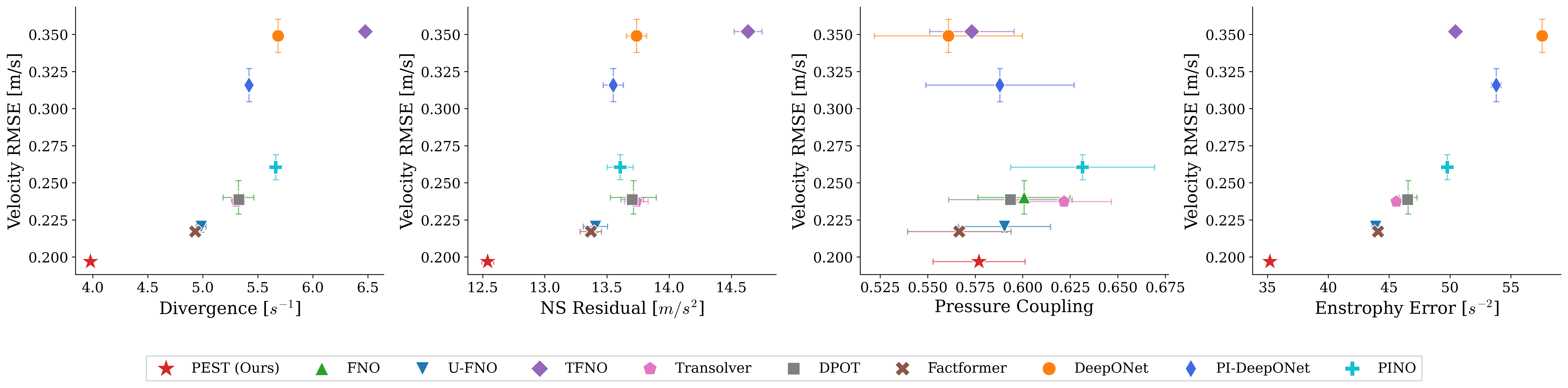}
\caption{
\textbf{Physical consistency on JHU (Round~3).}
After 15 autoregressive timesteps, PEST still achieves the lowest velocity RMSE and the smallest physics residuals.
Several baselines (notably DeepONet, PI-DeepONet, and TFNO) exhibit noticeably larger physics residuals compared to earlier rounds, indicating progressive drift from physical solutions.
}
\Description{Scatter plots showing physics consistency metrics versus velocity RMSE on JHU dataset for Round 3. PEST maintains best performance across all metrics.}
\label{fig:jhu_physics_r3}
\end{figure*}

Fig.~\ref{fig:jhu_physics_r1} in the main text presents Round~1 results. Here we provide the complete analysis for Rounds~2 and~3 (Figs.~\ref{fig:jhu_physics_r2} and~\ref{fig:jhu_physics_r3}).

\paragraph{Key Observations.}
Three findings emerge from the extended analysis:
(1)~PEST maintains its position at the lower-left corner (lowest RMSE and lowest physics residuals) across \emph{all} metrics throughout R1--R3, demonstrating that the physics constraints provide \emph{sustained} regularization rather than diminishing with error accumulation.
(2)~PEST exhibits relatively small error bars across test samples, indicating robust and consistent predictions across different flow realizations.
(3)~Several baselines---notably DeepONet, PI-DeepONet, and TFNO---show increasing physics residuals in later rounds, suggesting that their predictions progressively drift from physical solutions.
In contrast, PEST's physics residuals grow more slowly, highlighting the regularization effect of explicit physics constraints.

\subsection{Taylor-Green Vortex Per-Timestep Analysis}
\label{app:tgv_timestep}

\begin{figure*}[t]
\centering
\includegraphics[width=\textwidth]{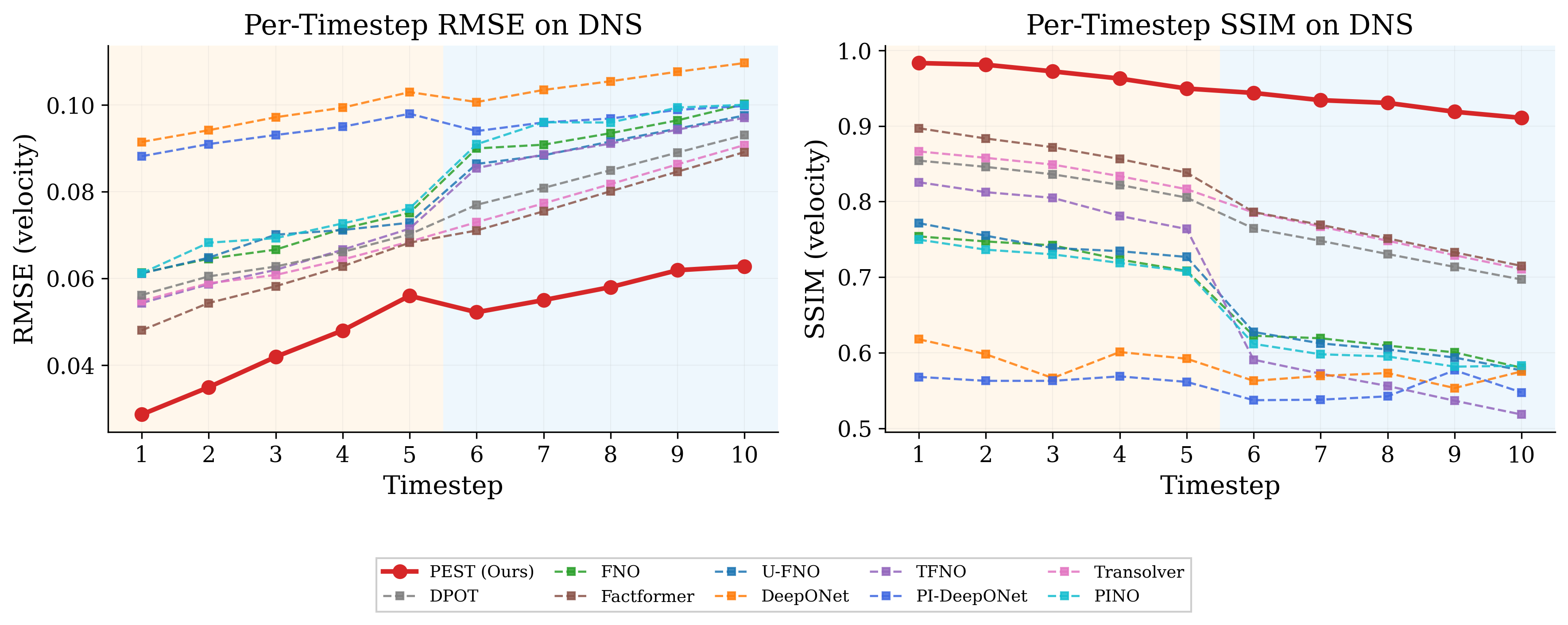}
\caption{
\textbf{Per-timestep RMSE (left) and SSIM (right) on Taylor-Green Vortex across 10 autoregressive steps (2 rounds).}
Timesteps 1--5 correspond to Round~1 and timesteps 6--10 to Round~2.
PEST (solid red) maintains SSIM above 0.9 for $u$, $v$, and $p$ components throughout the entire horizon, while all baselines degrade substantially in Round~2.
The widening gap in later timesteps demonstrates the benefit of physics-informed training for long-term stability under transient flow dynamics.
}
\Description{Line plots showing per-timestep RMSE and SSIM on Taylor-Green Vortex across 10 timesteps. PEST maintains the lowest RMSE and highest SSIM throughout.}
\label{fig:dns_rmse_ssim_comparison}
\end{figure*}

Fig.~\ref{fig:dns_rmse_ssim_comparison} shows per-timestep RMSE and SSIM on the Taylor-Green Vortex across the full 10-step prediction horizon (2 autoregressive rounds).

\paragraph{Takeaway.}
Unlike the statistically stationary JHU turbulence, the Taylor-Green Vortex exhibits transient dynamics spanning multiple regimes: initial vortex stretching, turbulent breakdown at peak dissipation ($t/t_c \approx 8$), and subsequent decay.
PEST maintains SSIM above 0.9 for $u$, $v$, and $p$ components throughout the 10-step horizon, indicating faithful preservation of flow structures even during the challenging transition-to-turbulence phase.
The performance gap between PEST and baselines widens in Round~2, confirming that physics-informed training provides increasingly important regularization as autoregressive errors accumulate.

\section{Detailed Spectral Ablation Analysis}
\label{app:spectral_ablation}

\begin{figure*}[t]
\centering
\includegraphics[width=\textwidth]{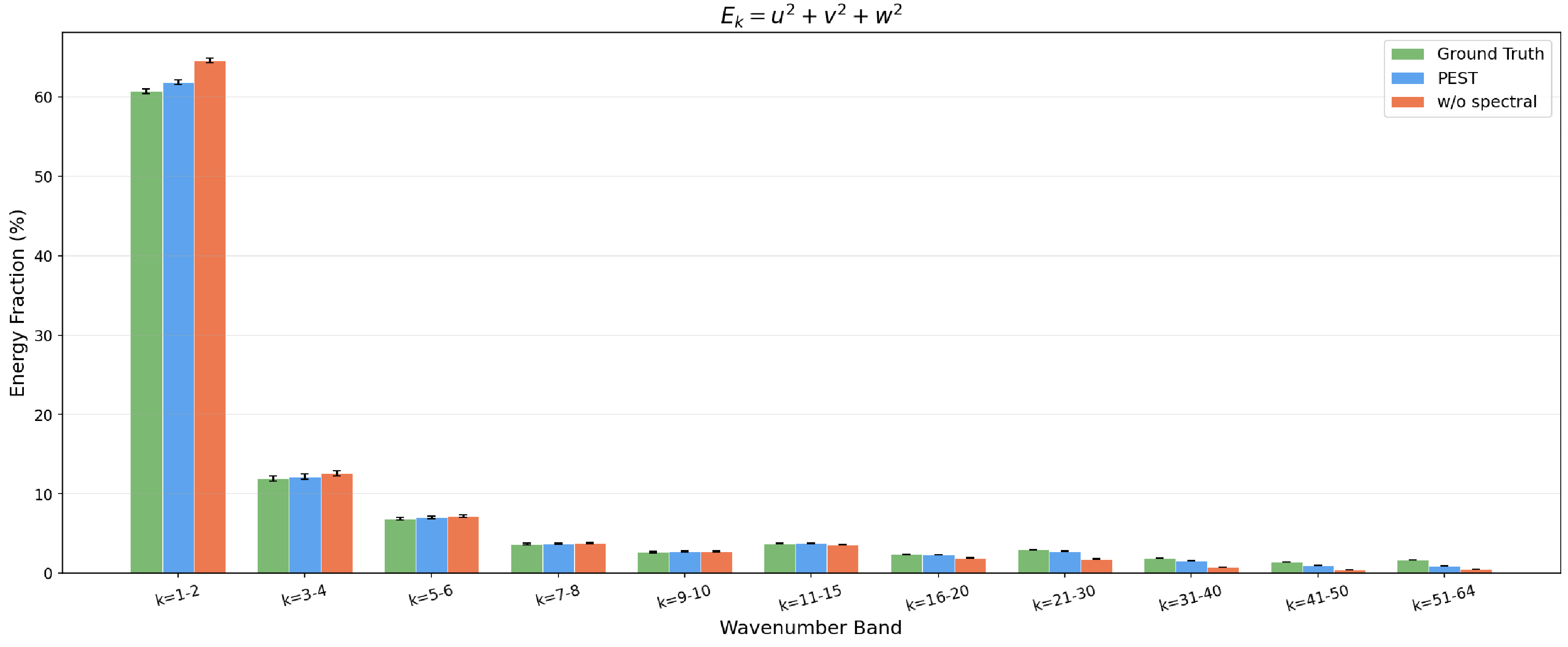}
\caption{
\textbf{Detailed spectral ablation on JHU Isotropic Turbulence.}
Kinetic energy fraction across 11 wavenumber bands for the DNS solution (green), PEST with spectral loss (blue), and PEST without spectral loss (red).
The frequency-adaptive spectral loss improves energy agreement with DNS across most bands---particularly at low ($k=1$--$2$), mid-to-high ($k \geq 11$), and high ($k > 20$) frequencies.
Only the $k=3$--$10$ range shows comparable performance with and without spectral weighting, as these bands already receive sufficient optimization emphasis from standard $\ell_2$ training.
}
\Description{Bar chart comparing energy distribution across 11 wavenumber bands for DNS solution, PEST with spectral loss, and PEST without spectral loss.}
\label{fig:spectral_ablation_detailed}
\end{figure*}

This section provides a fine-grained frequency decomposition that extends the coarse five-band spectral ablation presented in Fig.~\ref{fig:spectral_ablation} of the main text.
We partition the wavenumber range into 11 bands and compare the kinetic energy distribution ($E_k = u^2 + v^2 + w^2$) of the DNS solution against PEST predictions with and without the frequency-adaptive spectral loss (Fig.~\ref{fig:spectral_ablation_detailed}).

\paragraph{Takeaway.}
The fine-grained analysis reveals that the frequency-adaptive spectral loss provides broad-spectrum improvement rather than benefiting only a narrow frequency range.
The improvement spans both the lowest wavenumbers ($k=1$--$2$), where slight energy overshoot is corrected, and the mid-to-high wavenumbers ($k \geq 11$), where energy undershoot from standard $\ell_2$ training is mitigated.
This validates the physical motivation of our adaptive weighting: by explicitly rebalancing optimization emphasis across scales, the model achieves more faithful reproduction of the full turbulent energy cascade.

\section{Spectral Energy Characterization of Datasets}
\label{app:spectral_analysis}

\begin{figure*}[t]
\centering
\includegraphics[width=\textwidth]{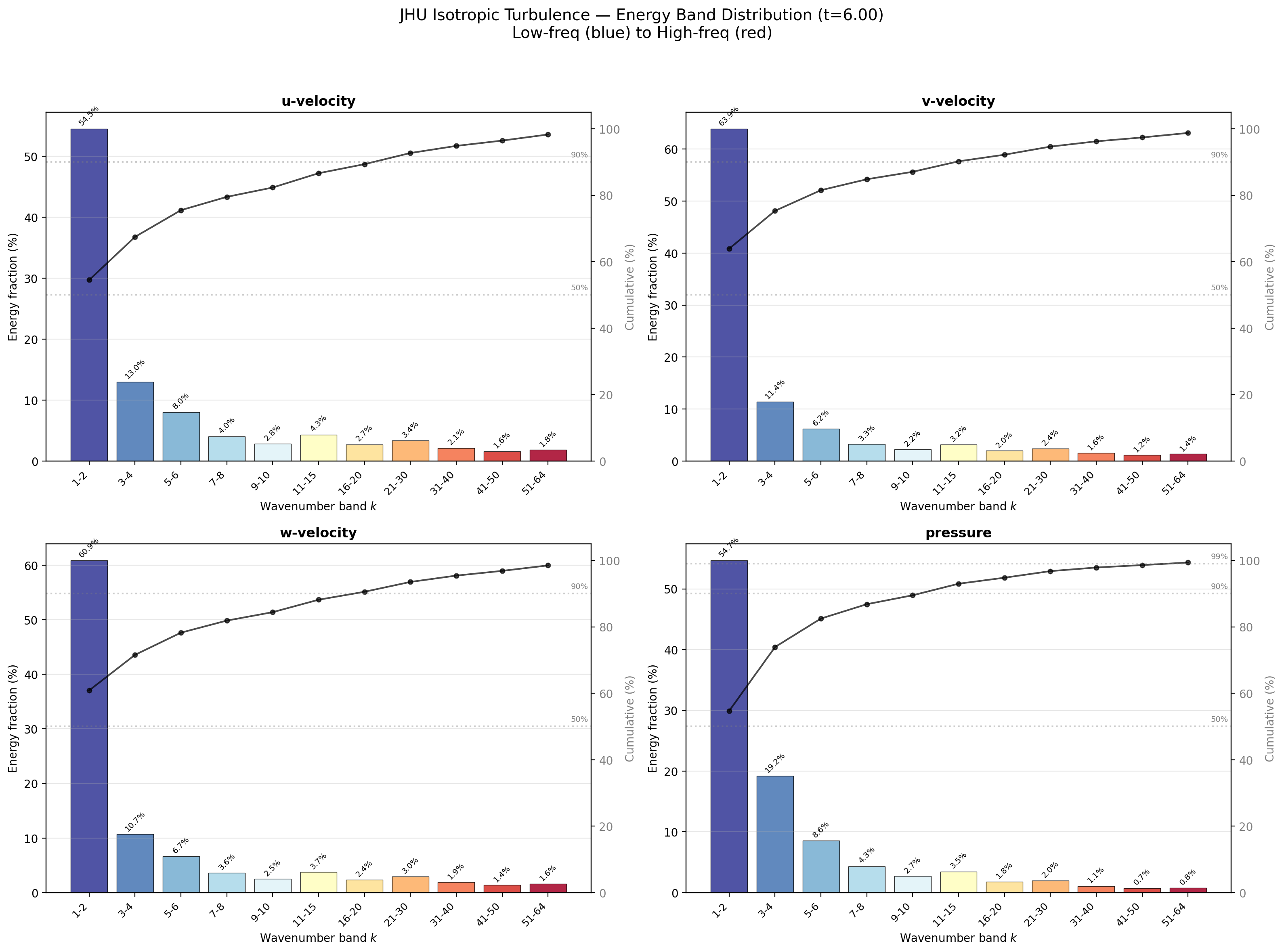}
\caption{
\textbf{Energy band distribution for JHU Isotropic Turbulence ($t=6.0$).}
Each panel shows the energy fraction (bars, left axis) and cumulative distribution (line, right axis) for one field component across 11 wavenumber bands.
The three velocity components ($u$, $v$, $w$) exhibit nearly identical distributions (statistical isotropy), with $\sim$30\% of energy residing beyond the lowest band and 3--5\% in high-frequency bands ($k > 30$).
This broad spectral support means that standard $\ell_2$ loss leaves significant high-frequency energy under-represented.
}
\Description{Four-panel figure showing energy distribution across wavenumber bands for u, v, w velocity and pressure in JHU turbulence.}
\label{fig:jhu_energy_analysis}
\end{figure*}

\begin{figure*}[t]
\centering
\includegraphics[width=\textwidth]{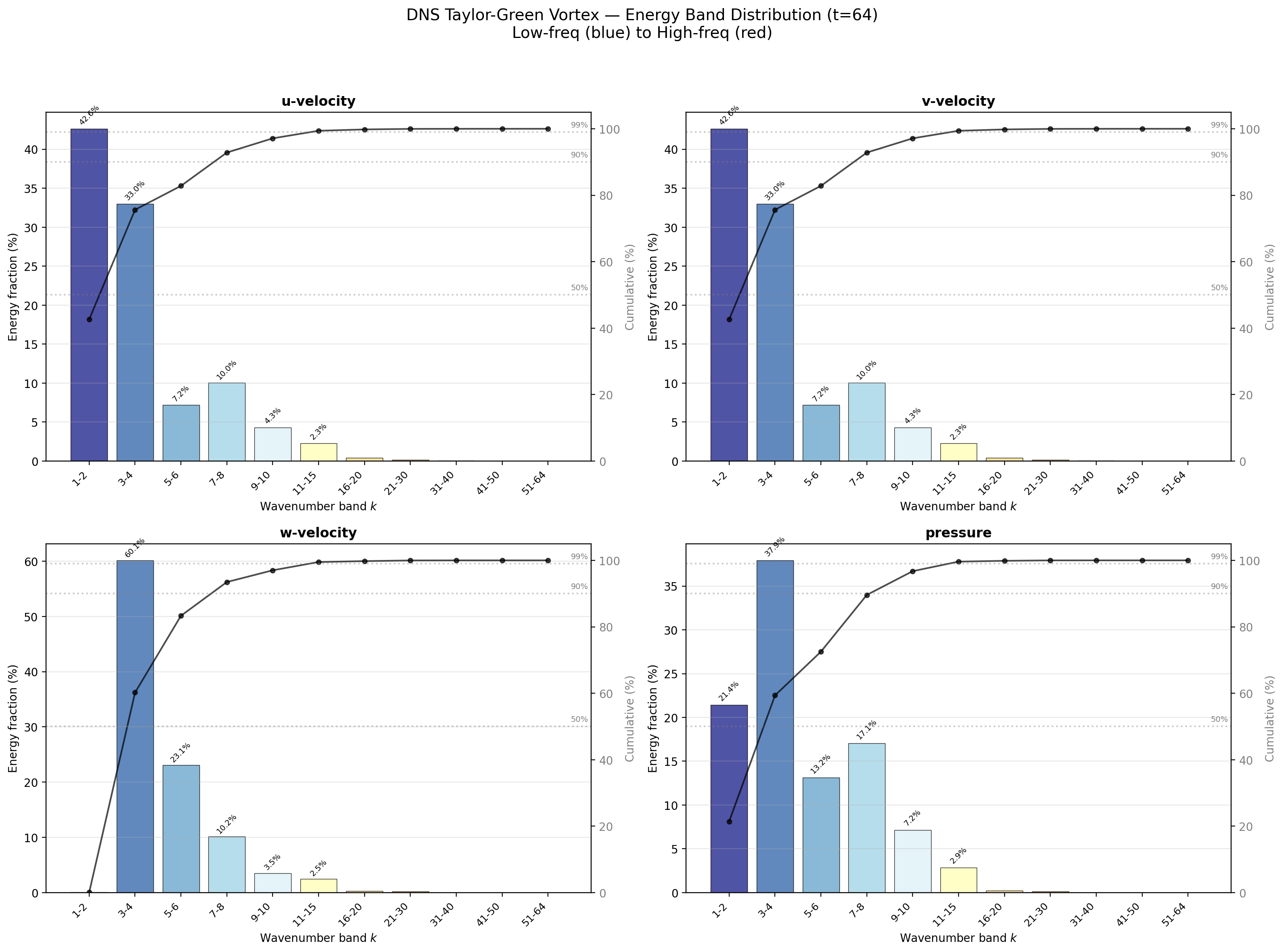}
\caption{
\textbf{Energy band distribution for Taylor-Green Vortex ($t=64$).}
In contrast to JHU, nearly all energy is concentrated below $k \approx 20$, with negligible content ($<0.5\%$) at higher wavenumbers---more than an order of magnitude less than JHU.
The $w$-component shows notable anisotropy compared to $u$ and $v$, and the pressure field has a distinct large-scale energy distribution.
}
\Description{Four-panel figure showing energy distribution for Taylor-Green Vortex. Components show anisotropy and minimal high-frequency energy.}
\label{fig:dns_energy_analysis}
\end{figure*}

To explain why the frequency-adaptive spectral loss yields more pronounced improvements on JHU than on Taylor-Green Vortex (\S\ref{sec:ablation}), we analyze the intrinsic energy distribution characteristics of both datasets (Figs.~\ref{fig:jhu_energy_analysis} and~\ref{fig:dns_energy_analysis}).

\paragraph{Takeaway.}
The contrasting spectral characteristics of the two datasets directly explain the ablation results.
JHU Isotropic Turbulence possesses broad spectral support with significant energy at high wavenumbers, making it highly susceptible to the scale bias of standard $\ell_2$ training---exactly the scenario where frequency-adaptive reweighting provides the greatest benefit.
Taylor-Green Vortex, with energy concentrated in a narrow low-frequency range, presents a smaller scale-imbalance problem, yielding more modest (though still positive) gains from spectral reweighting.
This analysis validates the physical motivation behind our frequency-adaptive loss design: the benefit scales with the degree of multi-scale energy distribution in the target flow.

\section{Plug-in Refinement Methods}
\label{app:plugin}

Beyond the core PEST framework, we evaluate two orthogonal refinement techniques that can be applied as model-agnostic plug-in components to most neural PDE solvers.
These methods are compatible with all our baseline models and ablation variants; we evaluate them on top of the full PEST model to assess potential for further improvement.

\subsection{Refinement with Low-Resolution Supervision}
\label{app:low_res_refine}

In computational fluid dynamics, high-resolution DNS is often prohibitively expensive, while coarse-resolution DNS or Large Eddy Simulation (LES) data can be obtained at significantly reduced cost.
Following~\cite{chen2024srtr}, we perform gradient-based refinement at test time: given low-resolution data $\mathbf{s}^{\text{LR}}$, we downsample predictions and minimize the degradation loss:
\begin{equation}
\mathcal{L}_{\text{deg}} = \| \text{Downsample}(\hat{\mathbf{s}}) - \mathbf{s}^{\text{LR}} \|_2^2.
\end{equation}

\subsection{PDE-Refiner: Denoising-Based Iterative Refinement}
\label{app:pde_refiner}

PDE-Refiner~\cite{lippe2024pderefiner} adapts the iterative refinement paradigm of diffusion models to neural PDE solvers.
During training, Gaussian noise of varying levels is added to DNS solution trajectories, and the model learns to predict the clean target:
\begin{equation}
\mathcal{L}_{\text{refiner}} = \mathbb{E}_{\epsilon, t} \left[ \| f_\theta(\mathbf{s} + \sigma_t \epsilon) - \mathbf{s} \|_2^2 \right],
\end{equation}
where $\sigma_t$ is a noise schedule and $\epsilon \sim \mathcal{N}(0, \mathbf{I})$.
At inference, the model iteratively refines its output through multiple denoising steps, progressively correcting accumulated prediction errors.

\subsection{Results}

\begin{table}[t]
\caption{Performance with plug-in refinement methods on JHU: RMSE ($\downarrow$). Both methods provide consistent improvements, with gains most pronounced in later rounds where accumulated errors are larger.}
\label{tab:sparse_gt_rmse}
\setlength{\tabcolsep}{2pt}
\footnotesize
\begin{tabular}{l ccc ccc ccc ccc}
\toprule
& \multicolumn{3}{c}{$u$} & \multicolumn{3}{c}{\cellcolor{bandgray}$v$} & \multicolumn{3}{c}{$w$} & \multicolumn{3}{c}{\cellcolor{bandgray}$p$} \\
Method & R1 & R2 & R3 & \cellcolor{bandgray}R1 & \cellcolor{bandgray}R2 & \cellcolor{bandgray}R3 & R1 & R2 & R3 & \cellcolor{bandgray}R1 & \cellcolor{bandgray}R2 & \cellcolor{bandgray}R3 \\
\midrule
PEST & .127 & .165 & .191 & \cellcolor{bandgray}.128 & \cellcolor{bandgray}.174 & \cellcolor{bandgray}.208 & .128 & .173 & .206 & \cellcolor{bandgray}\textbf{.060} & \cellcolor{bandgray}.086 & \cellcolor{bandgray}.104 \\
\addlinespace[0.5em]
+ PDE-Refiner & .128 & .163 & .186 & \cellcolor{bandgray}.129 & \cellcolor{bandgray}.171 & \cellcolor{bandgray}.202 & .130 & .170 & .200 & \cellcolor{bandgray}.061 & \cellcolor{bandgray}.085 & \cellcolor{bandgray}.102 \\
+ Sparse GT & \textbf{.126} & \textbf{.160} & \textbf{.182} & \cellcolor{bandgray}\textbf{.127} & \cellcolor{bandgray}\textbf{.168} & \cellcolor{bandgray}\textbf{.197} & \textbf{.128} & \textbf{.167} & \textbf{.195} & \cellcolor{bandgray}.061 & \cellcolor{bandgray}\textbf{.083} & \cellcolor{bandgray}\textbf{.098} \\
\bottomrule
\end{tabular}
\end{table}

\begin{table}[t]
\caption{Performance with plug-in refinement methods on JHU: SSIM ($\uparrow$). Both methods improve structural preservation, especially in later rollout rounds.}
\label{tab:sparse_gt_ssim}
\setlength{\tabcolsep}{2pt}
\footnotesize
\begin{tabular}{l ccc ccc ccc ccc}
\toprule
& \multicolumn{3}{c}{$u$} & \multicolumn{3}{c}{\cellcolor{bandgray}$v$} & \multicolumn{3}{c}{$w$} & \multicolumn{3}{c}{\cellcolor{bandgray}$p$} \\
Method & R1 & R2 & R3 & \cellcolor{bandgray}R1 & \cellcolor{bandgray}R2 & \cellcolor{bandgray}R3 & R1 & R2 & R3 & \cellcolor{bandgray}R1 & \cellcolor{bandgray}R2 & \cellcolor{bandgray}R3 \\
\midrule
PEST & .878 & .796 & .745 & \cellcolor{bandgray}.880 & \cellcolor{bandgray}.779 & \cellcolor{bandgray}.715 & \textbf{.873} & .774 & .708 & \cellcolor{bandgray}\textbf{.922} & \cellcolor{bandgray}.905 & \cellcolor{bandgray}.881 \\
\addlinespace[0.5em]
+ PDE-Refiner & .877 & .802 & .756 & \cellcolor{bandgray}.879 & \cellcolor{bandgray}.786 & \cellcolor{bandgray}.726 & .870 & .781 & .721 & \cellcolor{bandgray}.921 & \cellcolor{bandgray}.908 & \cellcolor{bandgray}.887 \\
+ Sparse GT & \textbf{.879} & \textbf{.808} & \textbf{.763} & \cellcolor{bandgray}\textbf{.881} & \cellcolor{bandgray}\textbf{.792} & \cellcolor{bandgray}\textbf{.735} & \textbf{.873} & \textbf{.787} & \textbf{.730} & \cellcolor{bandgray}.920 & \cellcolor{bandgray}\textbf{.912} & \cellcolor{bandgray}\textbf{.893} \\
\bottomrule
\end{tabular}
\end{table}

Tables~\ref{tab:sparse_gt_rmse} and~\ref{tab:sparse_gt_ssim} compare the two refinement strategies against the base PEST model on the JHU dataset.

\paragraph{Takeaway.}
Both refinement strategies provide consistent improvements, with gains most pronounced in later rounds (R2, R3) where accumulated autoregressive errors are larger.
Test-time refinement with sparse low-resolution supervision (+ Sparse GT) yields the largest improvement, reducing R3 RMSE by up to 5\% and increasing SSIM by up to 2 percentage points, by anchoring large-scale structures to external reference data.
PDE-Refiner provides a complementary, data-free improvement by using learned denoising to correct accumulated errors.
Importantly, both methods are general-purpose plug-in components applicable to any neural PDE solver, and their benefits are additive to the gains already achieved by PEST's physics-enhanced training.

\section{Additional Visualizations}
\label{app:visualizations}

This section provides comprehensive qualitative comparisons organized into two parts:
(1)~side-by-side comparisons of PEST against all baseline methods for each field component at the final autoregressive round, and
(2)~complete PEST rollout sequences across all rounds alongside the DNS solution to demonstrate long-term prediction stability.
Color bars in all figures indicate physical-space values (velocity in m/s, pressure in Pa) after denormalization.

\subsection{Comparison with Baseline Methods}

We present qualitative comparisons for all four field components ($u$, $v$, $w$, $p$) at the final autoregressive round for both datasets.
For JHU Isotropic Turbulence, we show Round~3 predictions ($t=6.0$); for Taylor-Green Vortex, we show Round~2 predictions.
In all comparisons, PEST predictions most closely match the DNS solution, preserving fine-scale turbulent structures that baselines fail to capture due to over-smoothing or structural degradation.

\subsubsection{JHU Isotropic Turbulence --- Round 3}

\begin{figure*}[t]
\centering
\includegraphics[width=\textwidth]{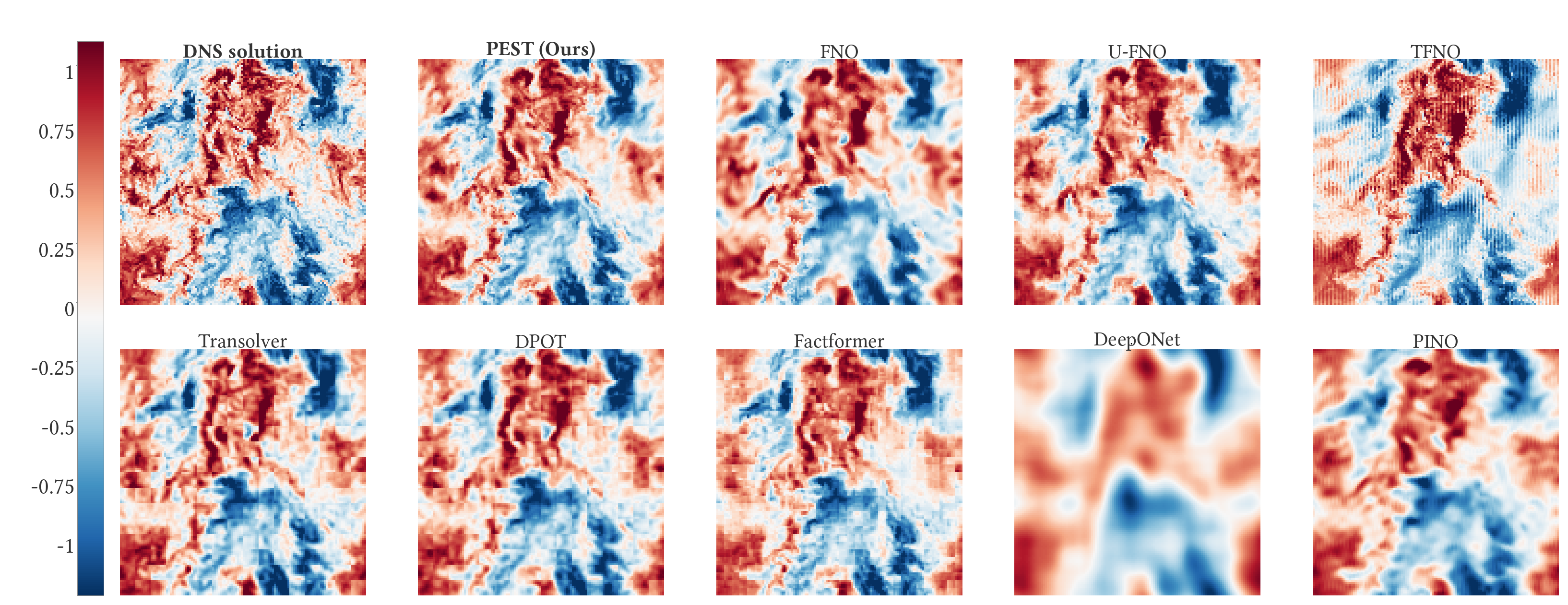}
\caption{\textbf{JHU $u$-component at autoregressive Round~3 ($t=6.0$).} PEST preserves both large-scale flow organization and small-scale fluctuations, whereas most baselines exhibit visible smoothing or structural distortion.}
\label{fig:jhu_u_round3_comparison}
\end{figure*}

\begin{figure*}[t]
\centering
\includegraphics[width=\textwidth]{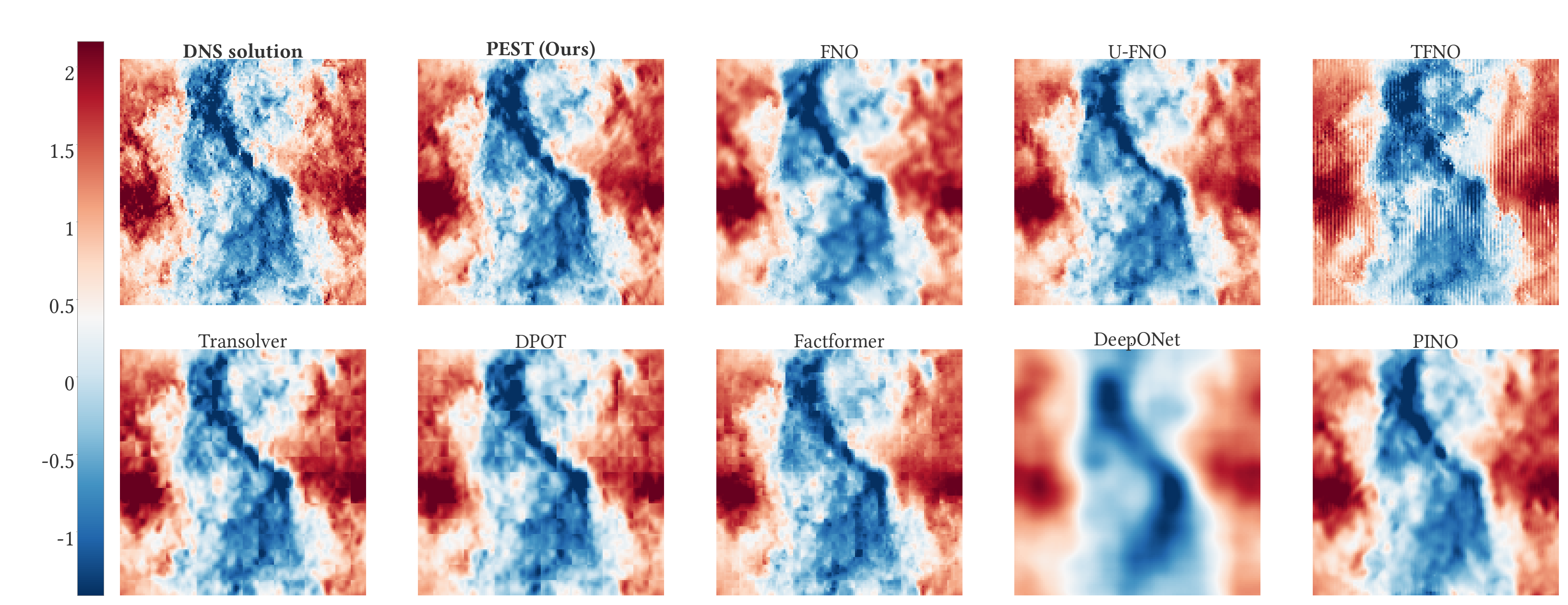}
\caption{\textbf{JHU $v$-component at autoregressive Round~3 ($t=6.0$).} Minimal degradation is observed at this advanced rollout stage, with PEST maintaining the highest fidelity among all methods.}
\label{fig:jhu_v_round3_comparison}
\end{figure*}

\begin{figure*}[t]
\centering
\includegraphics[width=\textwidth]{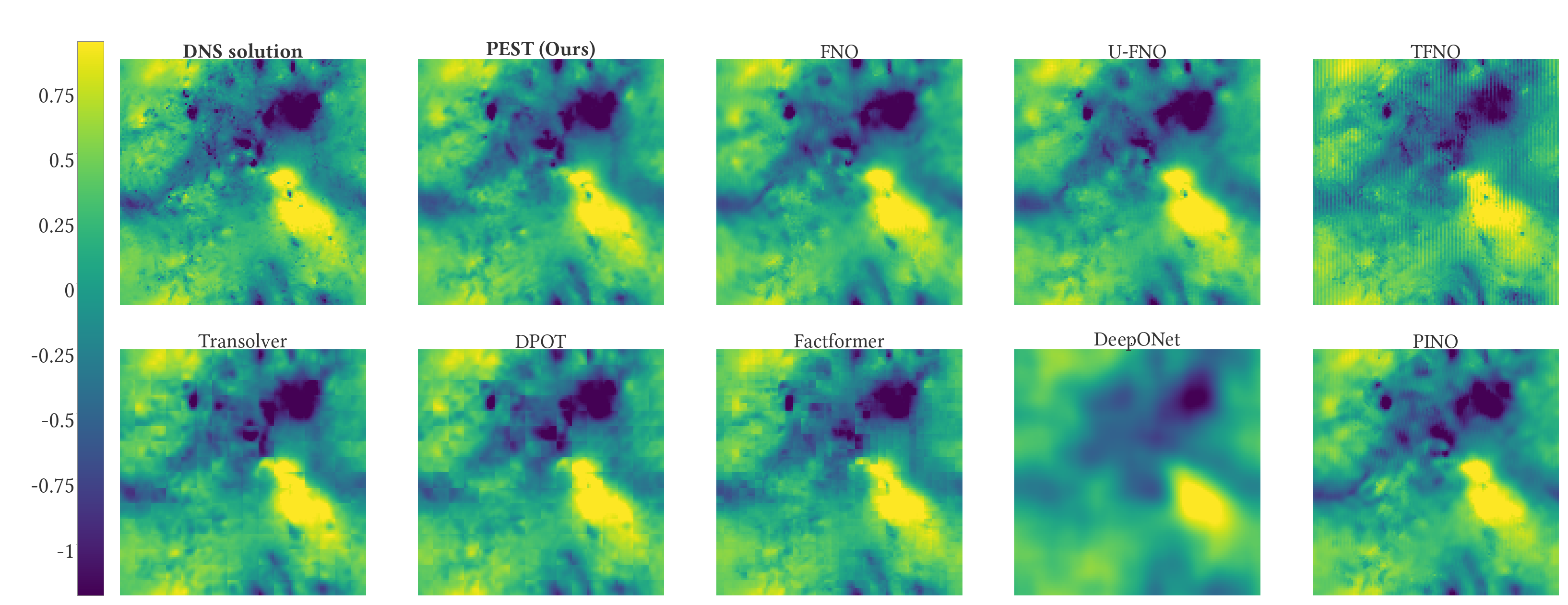}
\caption{\textbf{JHU pressure field at autoregressive Round~3 ($t=6.0$).} PEST accurately reproduces pressure fluctuations, demonstrating strong velocity--pressure coupling. Several baselines (notably DeepONet) show substantial deviation from the DNS solution.}
\label{fig:jhu_p_round3_comparison}
\end{figure*}

The $w$-component is presented in the main text (Fig.~\ref{fig:jhu_w_round3_comparison}). The remaining components are shown in Figs.~\ref{fig:jhu_u_round3_comparison} ($u$),~\ref{fig:jhu_v_round3_comparison} ($v$), and~\ref{fig:jhu_p_round3_comparison} ($p$).

\subsubsection{Taylor-Green Vortex --- Round 2}

\begin{figure*}[t]
\centering
\includegraphics[width=\textwidth]{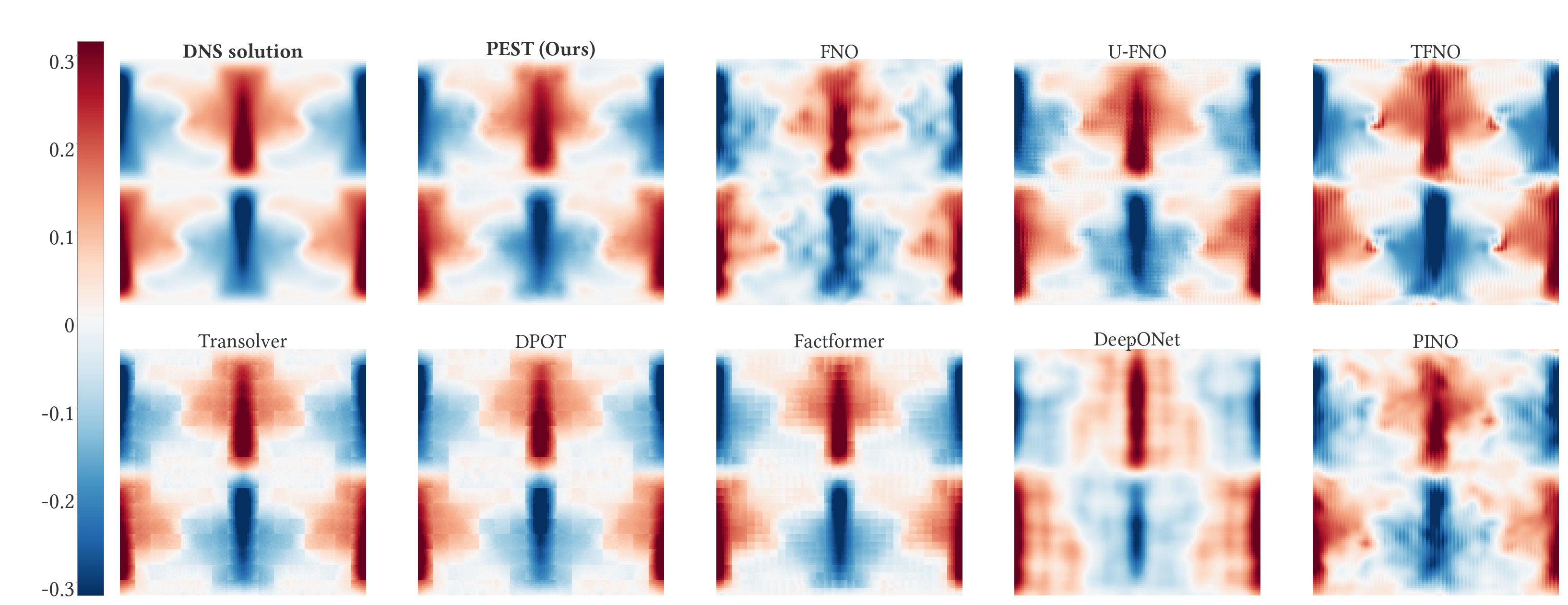}
\caption{\textbf{Taylor-Green $u$-component at autoregressive Round~2.} PEST captures the evolving vortical structures despite transient dynamics spanning the laminar-to-turbulent transition.}
\label{fig:dns_u_round2_comparison}
\end{figure*}

\begin{figure*}[t]
\centering
\includegraphics[width=\textwidth]{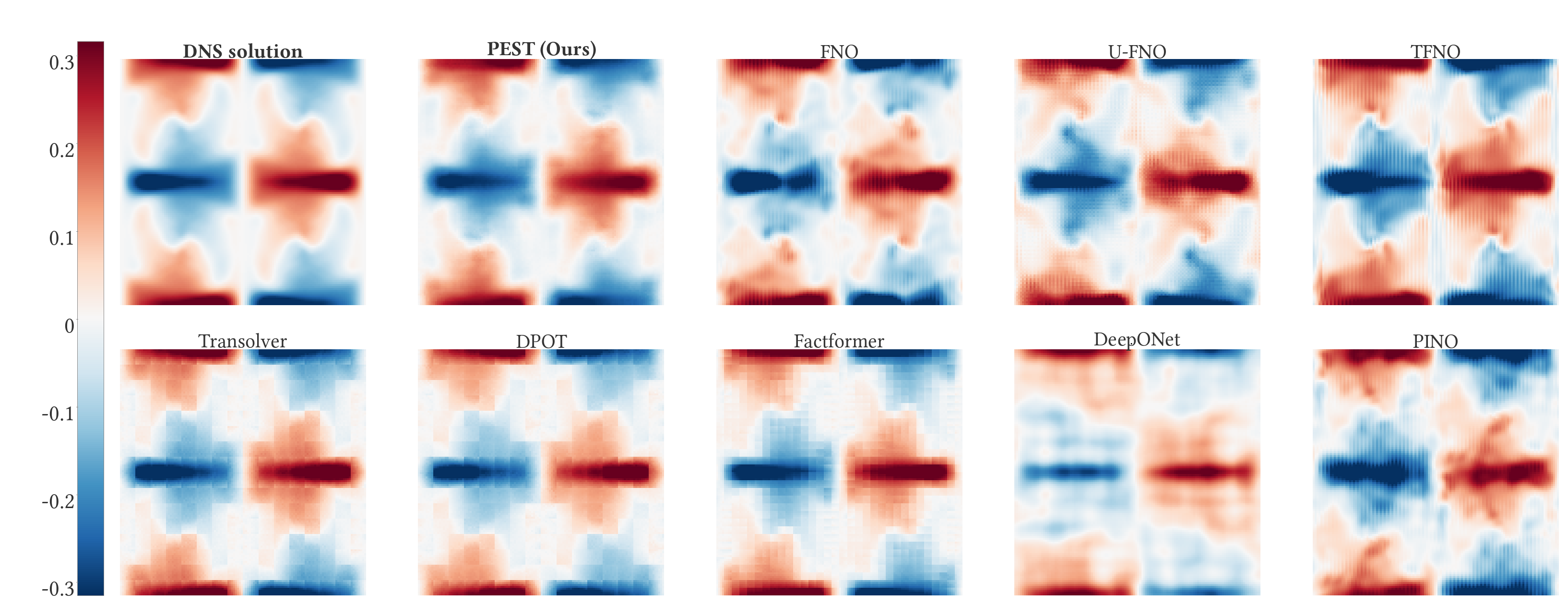}
\caption{\textbf{Taylor-Green $v$-component at autoregressive Round~2.} Structural fidelity is maintained during the turbulent transition phase, with PEST producing the closest match to the DNS solution.}
\label{fig:dns_v_round2_comparison}
\end{figure*}

\begin{figure*}[t]
\centering
\includegraphics[width=\textwidth]{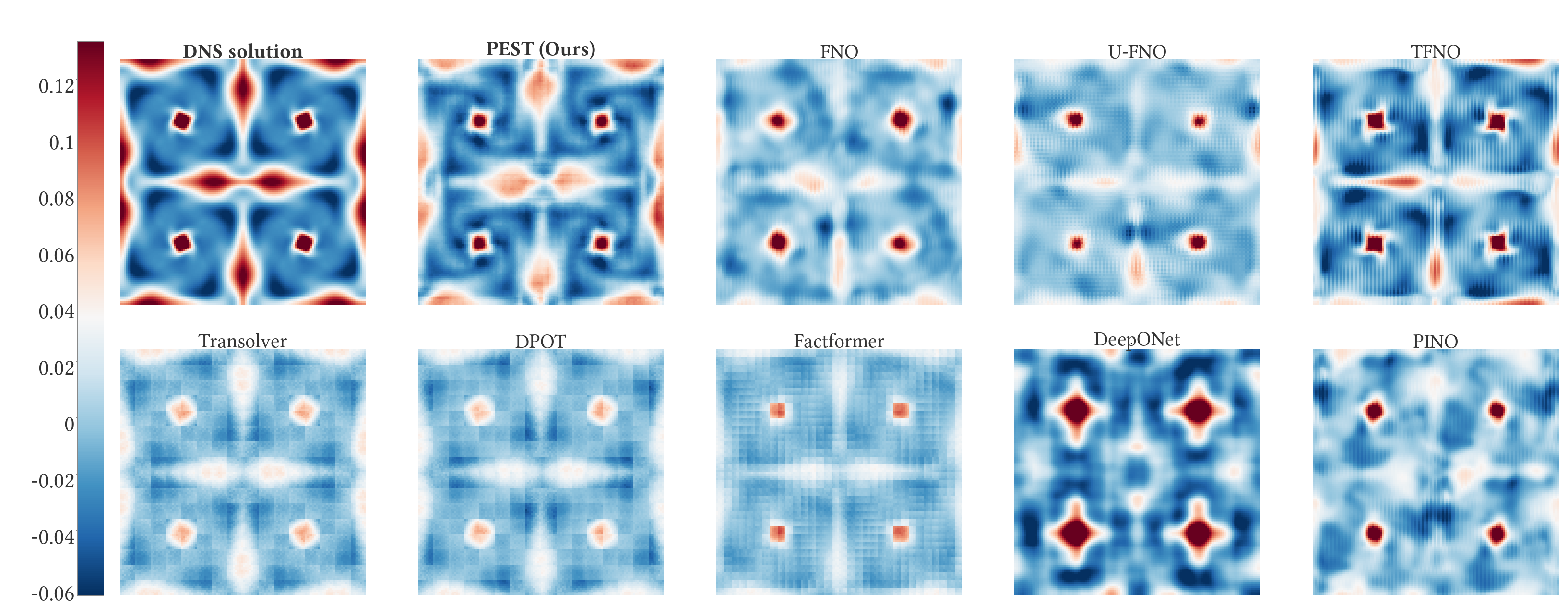}
\caption{\textbf{Taylor-Green $w$-component at autoregressive Round~2.} Fine-scale features are preserved by PEST, while baselines exhibit varying degrees of over-smoothing, particularly in regions of complex vortical interaction.}
\label{fig:dns_w_round2_comparison}
\end{figure*}

\begin{figure*}[t]
\centering
\includegraphics[width=\textwidth]{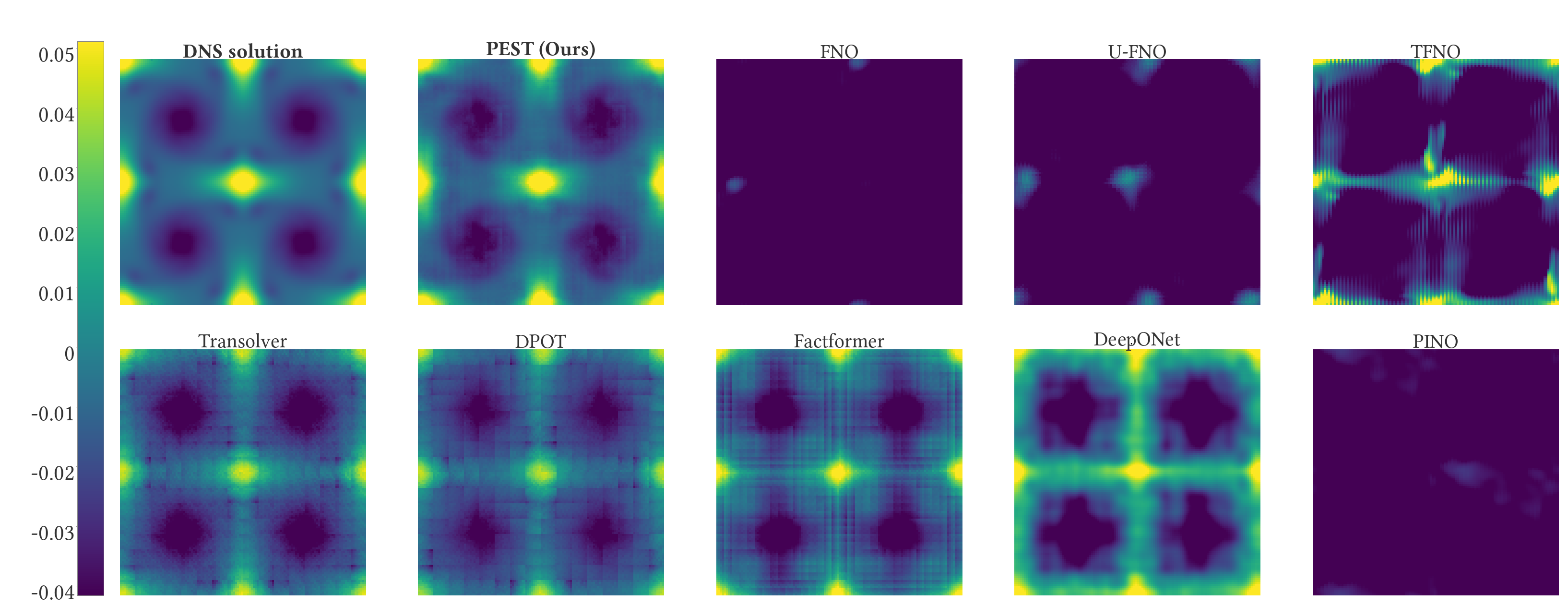}
\caption{\textbf{Taylor-Green pressure field at autoregressive Round~2.} PEST produces accurate pressure predictions consistent with the velocity dynamics, confirming that the learned velocity--pressure coupling generalizes to transient flow regimes.}
\label{fig:dns_p_round2_comparison}
\end{figure*}

Figs.~\ref{fig:dns_u_round2_comparison} ($u$),~\ref{fig:dns_v_round2_comparison} ($v$),~\ref{fig:dns_w_round2_comparison} ($w$), and~\ref{fig:dns_p_round2_comparison} ($p$) show predictions for the Taylor-Green Vortex at autoregressive Round~2.
The transient nature of this flow---undergoing vortex breakdown and turbulent transition---poses additional challenges beyond stationary turbulence, yet PEST maintains high fidelity throughout.

\subsection{Complete PEST Predictions Across All Rounds}

\begin{figure*}[t]
\centering
\includegraphics[width=\textwidth]{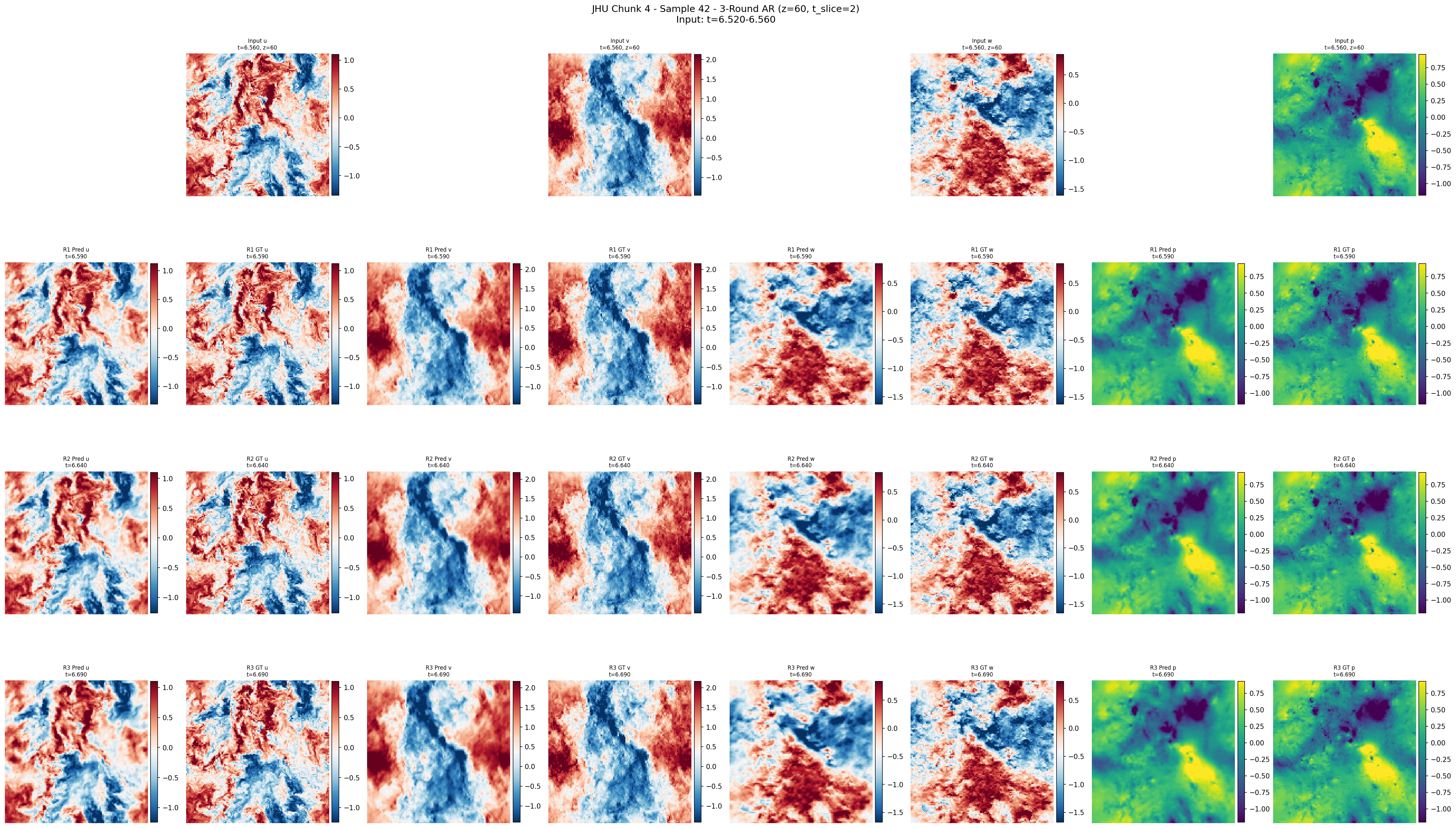}
\caption{\textbf{Complete PEST predictions on JHU across 3 autoregressive rounds (15 timesteps).} For each field component ($u$, $v$, $w$, $p$), PEST predictions are shown alongside the DNS ground truth across consecutive timesteps. PEST maintains visually accurate predictions throughout the entire 15-step rollout with minimal degradation, confirming the long-term stability observed in the quantitative metrics (Table~\ref{tab:jhu_main}).}
\label{fig:jhu_complete_pest}
\end{figure*}

\begin{figure*}[t]
\centering
\includegraphics[width=\textwidth]{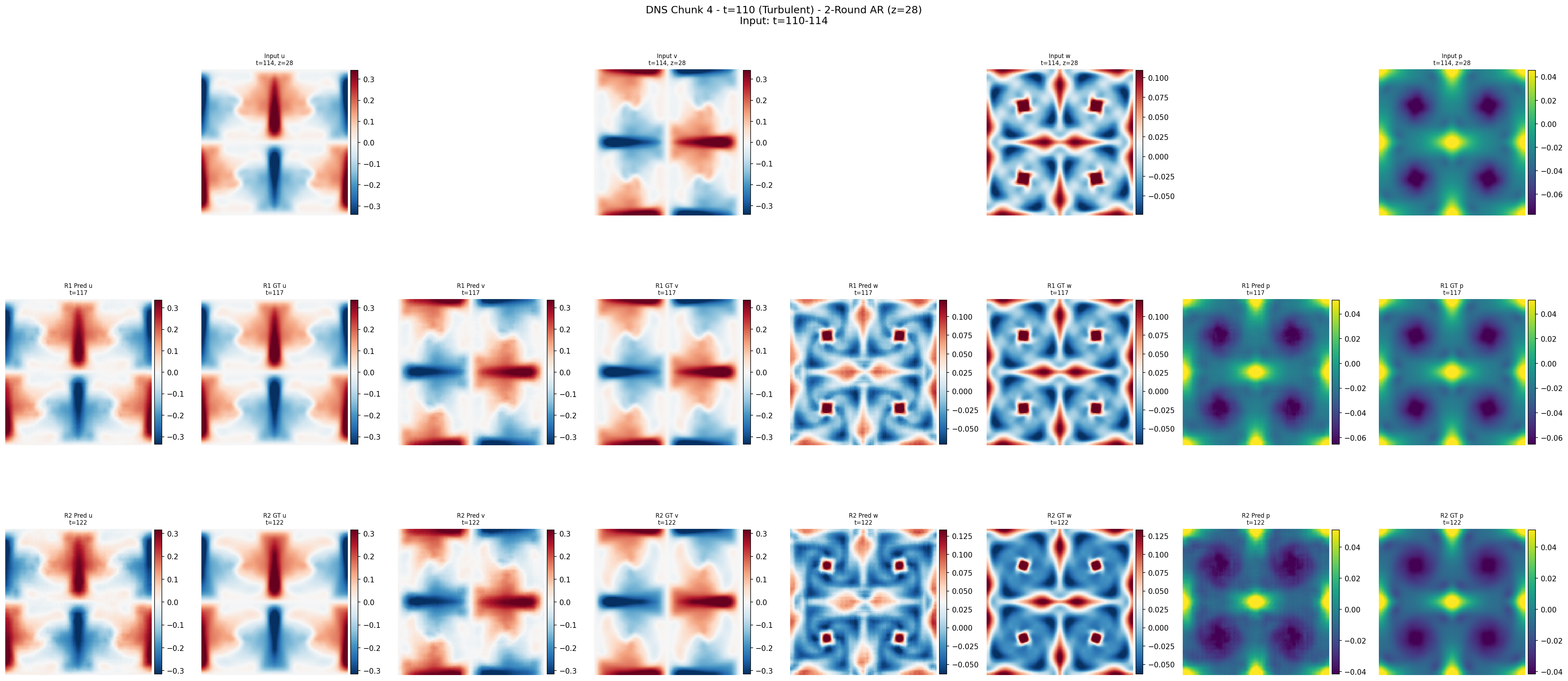}
\caption{\textbf{Complete PEST predictions on Taylor-Green Vortex across 2 autoregressive rounds (10 timesteps).} PEST faithfully tracks the flow evolution from organized vortical structures through turbulent breakdown, capturing the progressive development of small-scale features characteristic of the laminar-to-turbulent transition.}
\label{fig:dns_complete_pest}
\end{figure*}

To demonstrate the long-term stability of PEST, we show complete autoregressive rollout sequences with DNS solution comparisons for both datasets.
Each figure arranges the four field components ($u$, $v$, $w$, $p$) in groups, with PEST predictions paired alongside the corresponding DNS ground truth across consecutive timesteps.
Fig.~\ref{fig:jhu_complete_pest} presents the JHU rollout across 3 rounds (15 timesteps), and Fig.~\ref{fig:dns_complete_pest} presents the Taylor-Green Vortex rollout across 2 rounds (10 timesteps).

\end{document}